\documentclass[11pt]{article}

\usepackage{amsmath, amsfonts,amssymb, amsthm, euscript,makeidx,color,mathrsfs,latexsym,bm}
\usepackage{endnotes}
\usepackage{graphicx}
\usepackage{enumitem}
\usepackage{dsfont}
\usepackage{comment}
\usepackage{url}

\usepackage{caption}
\usepackage{subcaption}

\usepackage{natbib}
           \makeatletter
           \renewcommand{\bibsection}{
           \begin{center}

\section*{\refname\@mkboth{\MakeUppercase{\refname}}
               {\MakeUppercase{\refname}}}
           \end{center}
           }
           \makeatother

\usepackage{tikz}
\usepackage{pgfplots}
\pgfplotsset{compat=1.18}

\usepackage{tikz}
\usetikzlibrary{arrows.meta}
\usepackage{subcaption}

\usepackage[margin=1in]{geometry}

\newtheorem{assumption}{Assumption}

\newcommand{\ba}{\begin{array}}
\newcommand{\ea}{\end{array}}
\newcommand{\be}{\begin{equation}}
\newcommand{\ee}{\end{equation}}
\newcommand{\bee}{\begin{equation*}}
\newcommand{\eee}{\end{equation*}}
\newcommand{\bea}{\begin{eqnarray}}
\newcommand{\eea}{\end{eqnarray}}
\newcommand{\beaa}{\begin{eqnarray*}}
\newcommand{\eeaa}{\end{eqnarray*}}

\def\dbI{\mathbb{I}}

\def\dbR{\mathbb{R}}

\def\db1{\mathbb{1}}

\def\a{\alpha}

\def\e{\varepsilon}

\def\k{\kappa}

\def\h{\widehat}
\def\D{\Delta}

\def\cA{{\cal A}}

\def\cC{{\cal C}}

\def\cE{{\cal E}}

\def\cM{{\cal M}}

\def\cO{{\cal O}}
\def\cP{{\cal P}}

\def\cS{{\cal S}}
\def\cT{{\cal T}}

\def\no{\noindent}

\def\ms{\medskip}
\def\bs{\bigskip}
\def\q{\quad}
\def\qq{\qquad}

\def\cds{\cdots}

\def\ol{\overline}

\newcommand{\basa}{\begin{assumption}}
\newcommand{\easa}{\end{assumption}}

\newcommand{\bas}{\begin{assum}}
\newcommand{\eas}{\end{assum}}

\def\h{\widehat}

\def\cds{\cdots}

\def\ol{\overline}
\def\ul{\underline}

\def\dis{\displaystyle}

\def\1{{\bf 1}}

\def\:{\!:\!}
\def\reff#1{{\rm(\ref{#1})}}

at 9pt

\newtheorem{lem}{Lemma}

\newtheorem{prop}{Proposition}

\newtheorem{eg}{Example}

\newtheorem{assum}{Assumption}

\stepcounter{equation}
\begin{document}

\newcommand{\TITLE}{Democratic Policy Decisions with Decentralized Promises Contingent on Vote Outcome}


\newgeometry{margin=0.9in, top=0in}
\begin{titlepage}
	\title{Unlocking Democratic Efficiency: How Coordinated Outcome-Contingent Promises Shape Decisions\thanks{An earlier version of this paper was titled "Democratic Policy Decisions with Decentralized Promises Contingent on Vote Outcomes." We thank the students from the UBC reading group on political economy and finance in the summer 2022, and the participants in the UBC VSE Micro lunch workshop, the USC colloquium, the Bachelier Congress, the Stockholm School of Economics conference in memory of Tomas Björk, ESEM 2023 (Barcelona), the BIRS workshop in Banff, the Chicago Conference on Stochastic Analysis and Financial Mathematics, The Game Theory Days (UM6P, Rabat),  The CUHK colloquium, the NYU Abu Dhabi Economic Seminar.  We thank  Agostino Capponi, William Cassidy, Alvin Chen, Lorenzo Garlappi, Matthew Jackson, Navin Kartik, Chad Kendall, Henri Klintebaeck, Rida Laraki, Antonin Mac\'e, Nicola Persico, Dean Spencer, Tristan Tomala and John Wooders for comments. Ali Lazrak gratefully acknowledges the support from The Social Sciences and Humanities Research Council of Canada. Jianfeng Zhang gratefully acknowledges the support from NSF grants DMS-1908665 and DMS-2205972.}} 
	\author{ Ali Lazrak\thanks{ University of British Columbia, Sauder School of Business, email:  ali.lazrak@sauder.ubc.ca} and Jianfeng Zhang\thanks{University of Southern California, Department of Mathematics, email: jianfenz@usc.edu}.
	}
	\date{\today}
	\maketitle
	\begin{abstract}
		\noindent  We consider a committee voting on whether to adopt a reform under a quota rule, where members differ in how much they value the reform—some supporting it, others opposing it. We examine how members can influence each other’s votes through coordinated non-negative transfer promises, made prior to voting and contingent on the vote outcome. In equilibrium, these transfers are structured to prevent any coalition from profitably deviating in a coordinated way, while minimizing total transfers. We provide a complete characterization of these “strong” equilibria and show that they exist, are indeterminate, efficient, and involve transfers from high- to low-utility members. Such transfers prevent opponents from swaying less enthusiastic supporters and may be directed not only to opponents but also to lukewarm supporters.\\
		\noindent\textbf{Keywords:} Transfer promises contingent on vote outcome, political failure, majority coercion, direct democracy, strong Nash equilibrium, total transfer minimization\\
		\noindent\textbf{JEL Codes:} D70\\
		\bigskip
	\end{abstract}
	\setcounter{page}{0}
	\thispagestyle{empty}
\end{titlepage}
\restoregeometry
\pagebreak \newpage


\vspace{-1.3cm}

\newpage

\section{Introduction}

\vspace{-0.3cm}

Voting in political elections is often regarded as a moral duty, and trading votes for money or favors can be seen as abhorrent. Similarly, voting rights in corporations are a crucial way for shareholders to express their opinions on decisions that impact their economic ownership. Decoupling voting rights from economic ownership may, therefore, undermine the ideal of allocative efficiency in capital markets. 
Despite these negative connotations, vote trading remains a widespread practice that can take various accepted forms within legal and societal norms. For example, legislative logrolling is a common practice where politicians exchange their votes on particular issues in return for votes on other issues. Recent evidence of logrolling based on personal connection has been reported in the US Senate (\cite{cohen2014friends}) and in the US Congress (\cite{battaglini2023logrolling}). Early evidence from the 1840s includes the behavior of members of the British Parliament during the “railway mania.”\endnote{
Railway companies had to petition the British Parliament for a Private Act allowing them to begin construction of their lines. During the railway mania, an early case of a technology bubble, substantial funds were drawn from optimistic investors, including Members of Parliament (MPs). Parliamentary rules prevent MPs from directly voting on private acts concerning
companies at arm’s length, aiming to safeguard against personal interests influencing the approval of projects. Even so, evidence suggests that vote trading occurred between MPs, prioritizing individual interests (see Esteves and
Geisler Mesevage [2021]).}
In the case of corporate votes, activist investors may borrow shares for a nominal fee and use them to vote in favor of their own private agendas. Although commonly used in practice, the normative properties of vote trading are not well understood.
On one hand, Vote trading  can improve efficiency by allowing voters to express their preferences more precisely than with a simple binary vote (\cite{buchanan1962calculus} and \cite{coleman1966possibility}). On the other hand, it can hinder efficiency by imposing negative externalities on others (\cite{downs1957economic}).
In a recent survey, \cite{casella2021does} highlighted that ``Given the prominence of vote trading in all groups' decision-making, it is very surprising how little we know and understand about it."

In this paper, we evaluate one form of vote trading, where prior to casting their votes, the members of a committee make coordinated transfer promises contingent on the vote outcome. The committee, whether it is a legislative body or the shareholder base of a corporation, decides whether to enact a reform or reject it and retain the status quo. In our analysis, the committee decision is based on a quota rule, such as simple majority, supermajority, or unanimity. Committee members are fully informed about all relevant information, yet they disagree on their preferred alternatives because they derive different utilities from the passage of the reform. We assume that the reform is socially optimal in that the sum of utilities for the reform net of that for the status quo is positive.  Before voting, however, members can freely make credible and enforceable transfer promises contingent on the committee's decision. The promises are unconstrained and involve coordination within coalitions of any size, ranging from pairs to the entire committee.  The promises alter the incentives to vote, but voters retain control of their voting rights and cast their votes to maximize self-interest.

The transfer promises from our model capture some elements of the corporate bankruptcy proceedings where creditors and sometimes shareholders are asked to vote on a proposed plan of reorganization or liquidation, as opposed to resolving the issues through court intervention. The wedge between the expected court rulings' terms and the proposal's terms represents promises of transfers between stakeholders, as we envision in our model. Our model also reflects a common practice in legislative bodies, such as the US Congress, where bills are frequently amended before a final vote. The final bill bundles the initial bill and the amendments, which we view as transfers between congressmen that are only received when the bill passes. For example, the Patient Protection and Affordable Care Act (commonly known as Obamacare) was passed by the US Congress in 2010. During the legislative process, the bill was amended in order to gain enough support from both Democrats and Republicans. Several concessions to moderate Democrats and Republicans were made, including removing a public option and scaling back the scope of the bill. More generally, the promises can represent favors (e.g., logrolling),  legislative amendments, terms of liquidation, monetary payments, or any commitment to certain future actions that increase the advantage to the recipients of the promises.

To evaluate the practice of promises contingent on vote outcome, we formulate a two-stage game that we now describe informally. In the second stage, committee members take the transfers as given and vote for or against the reform.
In the first stage, players {\it simultaneously} announce transfer promises functions based on the vote outcome. The transfer promises are non-negative, enforceable and credible.  Once the promises are made but before the voting occurs, we allow players to deviate in a coordinated manner. Committee members are given the opportunity to form {\it blocking coalitions}, enabling them to influence the vote outcome and  achieve a more favorable outcome for all members of the blocking coalition. When forming a blocking coalition and deviating, members can modify existing transfer promises by either increasing or decreasing them. However, they may only reduce these transfers if the recipients explicitly consent. A recipient may agree to a reduction if the coalition’s deviation yields a higher overall payoff than they would receive otherwise. Absent recipient consent, the coalition members may only increase existing transfer promises or introduce new ones. We therefore assume the presence of cooperative norms that prevent group members from antagonizing one another by unilaterally retracting transfer promises against the will of the recipients. Our political equilibria are based on a stability criterion: We prohibit the formation of blocking coalitions in the first-stage game. We call these equilibria {\it strong equilibria}, as they represent Strong Nash Equilibria \cite{aumann1959acceptable} of our first-stage game, with the caveat that deviations are restricted to be contingent on the vote outcome and to exclude unilateral retractions.

We finally make the additional assumption that the total transfers by all committee members are minimized. In the absence of unilateral retraction, group members cannot undo excessive transfers and it is thus natural to assume that coalition coordinate on minimizing the transfers. By minimizing the total transfers, the unmodeled transaction costs associated with them are reduced. For instance, in politics, limiting amendments to a bill helps avoid deviation from its original purpose, especially if many changes might displease constituents or donors. Similarly, in logrolling or corporate liquidation, promises of future transfers can create uncertain costs and benefits. Minimizing these promises reduces uncertainty and related costs, and in some contexts, such as 19th-century British parliament during the ``railways mania," it can also minimize the risk of detection. We refer to such equilibria as {\it Strong Minimal equilibria} or ``SM" equilibria.

Our first result is to provide a complete characterization of strong equilibria. The characterization shows that strong equilibria exist and achieve the social optimum, meaning the reform passes. The efficiency result is a natural outcome, given that the entire committee has the ability to deviate, which implies that any strong equilibrium, if it exists, must be efficient. A more surprising result is the existence of strong equilibria. In the absence of transaction costs and informational frictions, a ``Coasean intuition" suggests that committee members should be able to contract to achieve an efficient outcome (\cite{coase2013problem}). 
While it's natural to model stable contracts using the strong Nash equilibrium in environments where coordination among group members is feasible, the criteria for achieving a strong Nash equilibrium is known to be quite demanding. Their existence is not always guaranteed, meaning that the Coasean intuition may not necessarily hold.

A contribution of this paper is to hold the Coasean intuition up to scrutiny within the framework we develop. In doing so, we provide an instance in which strong Nash equilibria exist and can be visualized  (see Figure~\ref{Fig: Illustration}). Two key assumptions ensure existence: (1) the transfers are contingent on the vote outcome, and (2) cooperative norms preclude unilateral retraction of existing transfer promises. We further show that the assumption of cooperative norms is also necessary: without it, the existence of even standard Nash equilibria is jeopardized.

Our existence result suggests that transfer promises contingent on vote outcomes enhance efficiency when committee members can coordinate their actions and uphold cooperative norms. Importantly, our results also show that when these norms are violated, the political equilibrium becomes unpredictable, emphasizing their essential role in maintaining stability.

In our second set of results, we characterize  SM equilibria and demonstrate that they exist but are indeterminate. While the minimal total transfer assumption reduces the set of strong equilibria, it does not imply uniqueness. Indeterminacy arises from the multiple ways in which total transfers can be divided among the promisers and transfers distributed among the recipients. However, the minimality assumption ensures that all equilibria share common characteristics, allowing us to predict the amounts and directions of transfer promises.

Specifically, there exists a critical voter who supports the reform, such that in all SM equilibria, the promisers have (weakly) higher utilities than this voter, while the recipients of the promises have lower utilities. Additionally, in all SM equilibria, transfer promises flow from committee members with higher utilities to those with lower utilities. Promisers are always reform supporters, whereas recipients can be either reform opponents or supporters, depending on the distribution of utilities and the quota rule. Intuitively, some reform supporters may receive transfers to insulate them from enticements offered by reform opponents.

When the reform is defeated in the absence of transfer, all SM equilibria are in ``the reaching across the aisle" type where the promisers support the reform while the recipients of the promises oppose it. In these equilibria, reform supporters compensate opponents for the total disutility they experience if the reform passes. When committee members are more polarized, the total disutility of the reform opponents increases, leading to a higher equilibrium total transfer.
Recently, {\it earmarking}, which refers to the practice of legislators allocating funds for specific projects in their district or state, has been revived in the US Congress.\endnote{ https://www.science.org/content/article/congress-restores-spending-earmarks-rules-remove-odor} Given that earmarks provide lawmakers with incentives to vote for legislation they may not support otherwise, they can be broadly interpreted as transfers within the framework of our model. The observed increase in polarization within the US Congress, coupled with the restoration of earmarking, aligns with the predictions of our model.


In the alternative scenario where reform supporters possess sufficient voting power to enact the reform, blocking coalitions can exist where opponents of the reform entice reform supporters with the weakest utilities, with transfer promises contingent on defeating the reform. These promises can persuade weakly motivated reform supporters to switch stances.
When these blocking coalitions are present, transfer promises made by reform supporters with the highest utility become crucial. These ``higher order"  promises help prevent the reform from being defeated after such coalitions form. We also show that when the reform is defeated without transfers, SM equilibrium transfers can be of the ``circle the wagon" type. This happens in equilibria where reform supporters with strong utility make some transfers to other reform supporters with lower utility.In these equilibria, when a committee operates under majority rule, reform supporters who are close to the median voter receive transfers from the strongest reform proponents. An example of this dynamic can be seen in the U.S. Senate, which was closely divided between Republicans and Democrats. Senator Joe Manchin, a moderate Democrat from West Virginia, exemplified this situation. His support was frequently pivotal for advancing legislation.\endnote{For instance, his influence on the inflation reduction Act is described in
\url{https://www.politico.com/newsletters/playbook/2022/08/08/how-it-really-happened-the-inflation-reduction-act-00050279}.}


{\bf Related literature.} The paper's main contribution is to show that transfer promises based on voting outcomes can be formally assessed.
Our results intersect with three streams of literature.

First, our findings contribute to the literature on political failures (e.g., \cite{becker1958competition}, \cite{wittman1989democracies}) by providing a formal mechanism to improve efficiency in collective decisions in presence of  majority coercion. The majority coercion is an important problem in many voting environments where a majority can impose outcomes that disproportionately harm minorities unless constrained by credible mechanisms, raising questions about fairness and long-term social stability. Our paper is closest to \cite{jackson2005endogenous}, who show that inefficiencies arise when players engage in transfers before participating in games with externalities (see also \cite{ellingsen2016confining}). While they focus on equilibria robust to unilateral Nash deviations, we study equilibria robust to coalitional deviations, leading to different efficiency results. Our approach also fully characterizes equilibria, including predictions on transfers, and explicitly addresses majority coercion in voting games, which \cite{jackson2005endogenous} do not cover.

Second, we contribute to the literature on vote trading.
Mechanisms such as vote storing (\cite{casella2012storable}), quadratic voting (\cite{eguia2023efficiencycollectivedecisionmakingquadratic}), vote delegation (\cite{berinsky2025tracking}) and trading votes in centralised markets (\cite{philipson1996equilibrium} ) and in strategic market games (\cite{xefteris2017strategic}) have been shown to enhance efficiency. \cite{casella2019trading} also provide a positive perspective on vote trading in the context of sequential logrolling.
We extend this literature by examining a framework in which voters can coordinate their actions, similar to \cite{casella2019trading}, but with a focus on trading through simultaneous promises.  We view this addition as important because many pre-vote interactions in corporate votes, referenda, or political elections can be thought of as transfer promises contingent on the vote outcome. 

Finally, we contribute to the literature on promises for political decisions, e.g., \cite{myerson1993incentives}, \cite{groseclose1996buying}, \cite{dal2007bribing}, \cite{dekel2008vote}, \cite{harstad2008side}, and more recently \cite{chen2022sequential}, \cite{louis2024buying} and, \cite{domenech2024buying}. This literature has largely concentrated on models with political representatives and revolves around leaders vying for votes by making pledges or campaign promises.
Rather than examining this practice solely within the political agency framework, we demonstrate how pre-vote interactions through promises can facilitate agreement and drive efficient outcomes in direct democracies.  By exploring the potential benefits of vote-contingent promises in this distinct context, our research underscores the need for a broader examination of this practice.


The paper proceeds as follows. Section \ref{Section Model} sets forth the general setting, Section \ref{Section: Strong equilibria} defines our equilibria. Section~\ref{subsection: Alternative} discusses the assumptions underlying our equilibria. Section \ref{Section: Equilibrium characterization}  characterizes
our equilibria. Section \ref{section: SM equilibria} defines the Strong minimal (SM) equilibria and shows their existence, indeterminacy and general implications. We illustrate more detailed implications of SM equilibria in Section \ref{Section: Equilibrium promises} and discuss a dynamic implementation of SM equilibria in Section~\ref{Section: Discussion}.
We conclude in  Section \ref{Section: Conclusion}.
The appendix provides the proofs of the paper's propositions.  The supplemental web appendix presents additional generalizations related to the discussions in Sections \ref{Section: Equilibrium promises} and  \ref{Section: Discussion}.

\vspace{-0.3cm}

\section{The model} \label{Section Model}

\vspace{-0.3cm}

We consider a committee $\mathbb{I}=\{ 1,\cds ,I\}$ of $I$ members or players. The game has two stages.  In the first stage, players make decentralized promises of transfers. In the second stage, players take the transfers as given and engage in voting. We now describe in detail the game starting with  Stage $2$.


{\it Stage $2$: The voting game.} In the second stage game, a collective decision on whether to accept a reform or reject it in favor of the status quo is taken by the committee. Each player $i$ must vote for the reform ($v_i=1$) or against it ($v_i=0$). We consider the pure strategies space $X=\{0,1\}$ and the associated set of strategy profiles $X^I$. We denote by $v_i$ and $v=(v_1,..,v_I)$ generic elements of $X$ and $X^I$. The committee is ruled by a quota rule with threshold  $\k \in \mathbb{I}$ whereby the reform is enacted if at least $\k$ committee members vote in favor. 
 The threshold for adopting the status quo is denoted as $\h \k$ where
  $
 \h \k := I-\k +1.
$
The reform is defeated if the number of reform opponents is greater than or equal to $\h \k$. We denote the set of vote outcomes by $\cO = \{ R,S\}$, where $O=R$ (resp. $O=S$) indicates that the reform is adopted (resp. defeated). When the strategy profile $v$ is played, the vote outcome is given by $O=R$, when $\sum_i v_i \geq \k$ and, $O=S$, otherwise.

In the absence of transfers, player $i$'s utility is a function $U_i : X^I \rightarrow  \dbR $ defined by
\bea
\label{eq: definition of utility}
U_i(  { v} ) := \left\{ \ba{lll}
\dis u_i &\mbox{if} \q  \sum_i v_i \geq \k ;\\
\dis 0,\q &\mbox{otherwise};
\ea\right.
\eea
where we have normalized the (cardinal) utility derived from the status quo for each voter to $0$ and denote $u_i$  the utility experienced by voter $i$ when the reform is adopted. We order the utilities $u_1\leq \cds \leq u_I$ and denote the utility vector by $u=(u_1,\cds,u_I)$. Voters are divided in their stance on the reform and  we denote by  $n$ the number of reform opponents with $1\leq n<I$. We denote the coalition of reform opponents as $\cC^S := \{i: u_i < 0\} \equiv \{1,\cds, n\}$. The coalition of reform supporters is denoted by $ \cC^R := \{i: u_i \geq 0 \}\equiv \{n+1,\cds, I\}$. The specific rule favoring support for the reform in the event of a tie ($u_i = 0$) is not important for our conclusions.  From a utilitarian standpoint, we assume that the passage of the reform is efficient:
\bea
\label{equ: R is efficient}
\left.\ba{c}
\sum_{i\in \dbI} u_i > 0.
\ea\right.
\eea

 {\it Stage $1$: The transfer promises game.} In the first stage game, each  player $i$ announces a transfer function defined as a vector of functions $t_i = ( t_{i1},..., t_{iI}) $, where $t_{ij}: \cO \rightarrow \dbR^+$ represents the non-negative transfer function that player $i$ is promising to player $j$ conditional on the vote outcome,  with $t_{ii}=0$.
 Transfer promises are assumed to be credible and enforceable.  That is, any transfer promises contingent on a specific voting outcome \textit{must} be honored by the initiator of the promises, the ``promisers," in favor of the recipients, the  ``promisees," if that outcome is realized.  Importantly, promises of transfers aimed at a player cannot be refused by that player. However, in our equilibrium deviation, we allow transfer recipients to neutralize such transfers by making a reverse transfer to the promiser.

 We denote the set of admissible transfer promises $\cA:=\dbR^{I}_+ \times \dbR^{I}_+$ and the set of admissible transfer promises profiles $\cA^I$. Let $t_i$  and $t = (t_1,...,t_I)$ denote generic elements of $\cA$ and $\cA^I$, respectively. We denote the set of one-sided transfer promises contingent on adopting the reform as
 $\cA_R := \{ t_i \in \cA | ~t_i(S) = 0 \}$.  We use  $\cA_R^I$ to refer to the set of one-sided promises profiles.  
 Let  $t^0 = (t_1^0,...,t_I^0)$ represent the degenerate transfer defined by $ t_{ij}^0 (O) = 0$ for all  $O \in \cO$ and for all $i,j$.

  Given a transfer $t \in \cA^I$, the {\it net transfer} $r^t_i$ (resp. $s^t_i$) is the sum of all transfers promised to member $i$ by other committee members minus the transfers that member $i$ has promised to others contingent passing (resp. defeating) the reform. We have
\bea
\label{equ: Net transfer}
\left.\ba{c}
 r^t_i := \sum_{j} t_{ji}(R) - \sum_j t_{ij}(R),~~s^t_i := \sum_{j} t_{ji}(S) - \sum_j t_{ij}(S).
 \ea\right.
 \eea
 When $r^t_i > 0$, committee member $i$ is a net promisee, and she gets a net utility increase of $r^t_i$ when the reform is adopted by the committee. When $r^t_i <0$, committee member $i$ is a net promiser, and her utility decreases by $|r^t_i|$ when the reform is adopted by the committee. Symmetric results hold for the promises contingent on reform defeat $s_i^t$. Observe that, for any $t \in \cA^I$,  the net promises profiles $r^t=(r^t_1,\cds, r^t_I) $ and $s^t=(s^t_1,\cds, s^t_I) $ must belong to the budget set  $ \cP
 := \{ x \in \dbR^I \mid \sum_i x_i =0\}$.

 Transfer promises alter the incentives to vote for the reform, but voters retain their voting rights in the second stage game. If the promise profile $ t$ is in place and the voting profile $v$ is played, then the payoff to player $i$ is given by
\bea
\label{eq: payoff}
\pi_i(v,t) 
:=\left\{ \ba{lll}
\dis u_i + r^t_i&\mbox{ if } \q  \sum_i v_i \geq \k  ;\\
\dis s^t_i,\q &\mbox{ otherwise }.
\ea\right.
\eea

\section{Definitions of the equilibria} \label{Section: Strong equilibria}

Our definition of equilibrium is based on subgame perfection of the overall two stages game and strong Nash equilibria for the transfer game when coalitions can form and coordinate their actions. It is natural to define it by backward induction.

In the second stage voting game, players take the transfers as given and confront a binary collective decision ($R$ vs $S$). We assume sincere voting as a natural equilibrium: when voting on a binary issue with complete information, voting to maximize personal preferences constitutes an undominated Nash equilibrium.  Fixing the transfer $t \in \cA^I$ and denoting the indicator function $\1$ equal to one when a condition holds and zero otherwise, the equilibrium voting strategy of player $i$ in the second stage game is sincere and the corresponding vote outcome are
\bea
\label{eq: sincere voting}
 v_i(t) = \1_{u_i+r^t_i \geq s_i^t},\qq
O(t) = R~ \1_{\sum_i v_i(t) \geq \k } + S  ~\1_{\sum_i v_i(t) < \k }.
\eea

We now define the equilibrium in the overall two stages game associated to the transfer $t \in \cA^I$. Using sub-game perfection, the payoff profile is denoted by $\h \pi(t) = (\h \pi_1(t),\cds,\h\pi_I(t)) $, where $\h \pi_i(t) = \pi_i(v(t) , t)$. We recall that $v(t) = (v_1(t),\cds,v_I(t))$ where $v_i(t)$ is defined in equation \reff{eq: sincere voting} for each $i$. Therefore, the first stage game can be seen as a one period game with payoffs profile function $\h \pi: \cA^I \rightarrow \dbR^I$. We refer to $\h \pi_i(t)$ as the {\it post-transfer utility} of member $i$.

{\it Blocking coalitions.}  We say that a  nonempty coalition of committee members $\cC\subset \mathbb{I}$ {\it blocks} the transfer function $t \in \cA^I$ if there exists a transfer function $\tilde{t} \in \cA^I$ such that
\begin{enumerate}
\item  The coalition $\cC$ includes both the initiators of the deviations and the transfer recipients subject to retractions in the deviation:
\bea
\label{coalition-retraction}
 \cC =\{ i ~|~ \tilde t_i \neq t_i \} \cup  \{i~|~ \tilde t_{ji}(O) < t_{ji}(O) \mbox{ for some } j \mbox{ and } O \},
\eea
\item  The deviation benefits all members of $\cC$: $\h \pi_i(  \tilde t) > \h \pi_i(t)  $ for all $i \in \cC$.
 \end{enumerate}

Equation \reff{coalition-retraction} defines the blocking coalition to include both the members who initiate transfers and those whose pre-deviation transfers are reduced. By including the recipients of retracted transfers, the definition ensures that their consent is necessary—making them full participants in the coalition formation process. In contrast, members whose received transfers increase are not included in the blocking coalition, since they can offset the increase by initiating reverse transfers if they so choose. Such neutralizing actions are not available to members subject to retractions, as negative transfers are not permitted. This asymmetry justifies their inclusion in the coalition. A key implication of this definition is that, if a blocking coalition exists, it necessarily overturns the committee’s decision, as established in the following lemma:

\begin{lem} \label{lem: BC overturn}
If the coalition $\cC$ is blocking the transfer $t \in \cA^I$ with the deviation $\tilde t \in \cA^I$, then the deviation overturns the committee decision, that is, $O(\tilde t) \neq O(t)$.
\end{lem}

The Lemma shows that overturning the committee's decision is a result of the definition of blocking coalitions, {\it not} a defining characteristic of them.
 A {\it Strong sub-game perfect  equilibrium} is a transfer function $t \in \cA$ such that there exists no coalition $\cC$ that can block it. We denote the set of these equilibria by ${\pmb \cS}$. In our discussions, we refer to these as {\it strong equilibria}.

\vspace{-0.3cm}

\section{Discussion of the model's assumption} \label{subsection: Alternative}

\vspace{-0.3cm}


When merging the two stages, our game can be thought of as an endogenous agenda-setting model in a voting game with a multidimensional infinite set of alternatives, that is the set of all transfers. A key contribution of this paper is to establish the existence of the set $\pmb \cS$ (see Section~\ref{Section: Equilibrium characterization}).  This existence result stands thus in sharp contrast to the findings of the ``chaos theorems" literature, which emphasize the non-existence of stable equilibria in voting games with multidimensional alternatives (See, e.g., \cite{plott1967notion}, \cite{mckelvey1976voting}, \cite{schofield1983generic} and \cite{duggan2000strategic}). Besides technical reasons,\endnote{Unlike the chaos theorem literature, which assumes a compact set of alternatives, our set $\cP$ is unbounded and lacks a bliss point. Thus, transfers lack the Condorcet winner property, as a binary vote can always be swayed by forcing one member to promise large transfers to all other members. Moreover, discontinuous preferences in our framework render Plott’s gradient method inapplicable, as it relies on first-order conditions for equilibrium.}  this result also relies on two key assumptions: first, that transfers are contingent on the voting outcome; and second, that blocking coalitions are restricted from antagonizing others through unilateral retractions. We now discuss the significance of these assumptions and their motivation.

{\it Transfer contingent on vote outcome.} This assumption reflects processes like amending political bills or resolving corporate bankruptcy. It also captures aspects of clientelism, where politicians (“patrons”) grant favors in exchange for electoral support (see, e.g., \cite{weingrod1968patrons}). Since favors are granted only if the politician is elected, conditioning transfers on vote outcomes is a reasonable assumption in both contexts.

More broadly, transfers could also be based on individual votes or entire voting profiles. In particular, transfers contingent on recipients’ votes—often involving upfront payments—reflect practices such as corporate vote trading. When transfers depend on recipients’ own votes, preliminary analysis suggests that the existence of strong equilibria is uncertain, and thus it remains unclear whether the Coasean intuition holds. In contrast, transfers based on vote outcomes  exhibit a ``contracting on contracts" feature (\cite{katz2006observable}).  Intuitively, the interdependence of transfer promises contingent on vote outcome ties the continuation utility of the promisers to that of the promises recipients and supports the existence of a strong equilibrium. Our findings on the success of transfers contingent on vote outcomes in promoting efficiency suggest the importance of building institutions that facilitate such transfers.
 Our results represent thus  a first step that lays the groundwork for a comprehensive analysis, providing a basis for validating the ``Coasean intuition"  in the context of more general form of vote trading or potentially offering evidence to the contrary.

{\it The no unilateral retractions assumption.} To illustrate the importance of this assumption, assume that we allow deviations to retract without the consent  of the recipients and consider the simpler setting  of Nash equilibria of the first stage game.

 A transfer $t\in\cA^I$ is a  Nash equilibrium if  no single player $i$ can unilaterally deviate by changing the transfers from $t_i$ to  $\tilde t_i \in \cA$ and improve her payoff, that is,
$$
  \h \pi_i( t )  < \h \pi_i(\tilde t_i, t_{-i} ),
$$
where the  transfer promises $t_{-i}$ denotes the individual transfers of all players except $i$, and  where $(\tilde t_i, t_{-i}) =(t_1,\cds,t_{i-1},~\tilde t_i, ~t_{i+1}, \cds, t_I) $.

The following result shows 
that the existence of Nash equilibria is jeopardized in our game.

\begin{prop}
\label{prop: Non existence of Nash}[\textbf{Non-existence of Nash equilibria}] Consider a committee consisting of three members with utilities $u = (-3, 2, 2)$ and operating under majority rule, $\k = 2$.
In this scenario, no  Nash equilibrium exists.
\end{prop}

Proposition \ref{prop: Non existence of Nash} suggests that making predictions about the political equilibrium becomes impossible is our setting: if there is no Nash equilibria, then there is no hope to get stable equilibria that are robust to coalitional deviations. Although the scope of the proposition is limited to an example,\endnote{More broadly, it can be proven that the only Nash equilibria with unilateral retractions involve off-equilibrium transfers, where $t(O(t)) = 0$. These equilibria are less interesting, as the transfers are never executed and merely represent empty promises.} its proof provides insight into why equilibria are undermined. When considering deviations, promisers perceive certain transfer recipients as non-pivotal and, consequently, retract their transfers to these recipients since these retractions do not change the vote outcome. This ``unraveling" mechanism is unsurprising, as the possibility of withdrawing transfers compromises their credibility and undoes the equilibria.  The mechanism is very similar to the one that destabilizes equilibria in competitive markets for votes: in such markets, non-pivotal voters offer their votes for sale at any positive price, increasing supply and putting downward pressure on prices, thereby preventing the equalization of supply and demand (see \cite{philipson1996equilibrium}).

Our results therefore apply to environments where norms and institutions foster cooperation and prevent committee members from antagonizing others through unilateral retractions. In such cooperative settings, transfers contingent on vote outcomes promote efficiency, in line with the Coasean intuition. Conversely, our results highlight that when cooperative norms are absent, even the existence of Nash equilibria is jeopardized, and a fortiori, stability is undermined.

While our model emphasizes the importance of cooperative norms in identifying and predicting political equilibria, it does not explain the origin of these norms, as they are assumed in our setting. However, a possible channel could create incentives for their development. If a committee votes on different issues and preferences rotate randomly depending on the issue, any member could find themselves in the minority at some point. As a result, resolving the political failure of majority coercion becomes important for all members. In this setting, cooperative norms, combined with the practice of transfers contingent on vote outcome, can benefit all committee members as an alternative to voting without transfers. The norms may represent an equilibrium in a pre-stage game where the committee adopts a norm of cooperation that prevents unilateral retractions with no consent. The idea that endogenous cooperation can emerge an equilibrium outcome in a model with repeated voting has been recently developed in \cite{mace2024repeated}. In their model the cooperation takes the form of implicit logrolling. In Section~\ref{Section: Discussion}, we will explore sequential implementations of our equilibria. Under the assumption that transfers occur sequentially, with cumulative transfers at each step forming the status quo for a new game, the no-retraction assumption becomes equivalent to the non-negative transfer assumption in each subgame.

\vspace{-0.3cm}
\section{Equilibrium characterization} \label{Section: Equilibrium characterization}

\vspace{-0.3cm}

 We now provide a characterization of the set of strong equilibria, $\pmb \cS$.

\vspace{-0.1cm}
 \begin{prop} \label{prop: Strong equilibria} [\textbf{Characterization of strong equilibria}]
The set $\pmb \cS$ is non-empty and efficient, $O(t)=R$ for all $t \in \pmb \cS$. Moreover,
\bea
\label{eq: Strong Nash Characterisation}
\left.\ba{c}
 \pmb \cS = \{ t \in \cA^I \mid \sum_{  \cC} u_i+r^t_i \geq s^t_i \mbox{ for all coalitions } |\cC| \geq \hat \k   \}.
 \ea\right.
 \eea
 \end{prop}
The characterization~\reff{eq: Strong Nash Characterisation} of Strong equilibria requires that once the transfers are implemented, the reform remains socially optimal for any coalition that has the power to overturn it. Strong equilibria are then characterized by the system of linear inequalities \reff{eq: Strong Nash Characterisation} on net transfers, forming a convex polyhedron within the set $(r,s) \in \cP^2$. The characterization also implies the monotonicity of the set ${\cal S}$ with respect to the quota rule: if the inequalities \reff{eq: Strong Nash Characterisation} hold for $\h \k$, they will also hold for any quota $\h \k'$ such that $\h \k' \geq \h \k$. This means that, for any quota $\k$,  strong equilibria under the quota rule $\k$ are also strong equilibria under quota rule $\k-1$.
In particular, the set of strong equilibria under the unanimity rule is a subset of the set of strong equilibria under the majority rule.

Proposition \ref{prop: Strong equilibria} shows that the reform passes in all strong equilibria. The Coasean intuition holds: By offering transfers before voting, supporters can influence opponents through promises contingent on the reform’s approval.

Notice that the system of inequalities \reff{eq: Strong Nash Characterisation} restricts the transfer functions $t$ solely through the net transfers $(r^t, s^t)$. The system has in general multiple solutions suggesting the indeterminacy of the strong equilibria.  Observe that if the net promises $(r,s) \in \cP^2$  satisfy the inequalities \reff{eq: Strong Nash Characterisation} then, by linearity, the one-sided net promises $(\h r= r-s, \h s =0)$ also satisfies those inequalities. Hence any one-sided transfer function $ \hat t \in \cA^I_R$  satisfying $r^{\h t} = \h r$ is also a strong equilibrium. This means that one-sided transfer promises contingent on reform adoption are sufficient to reach efficiency.  We discuss an example to illustrate the results of Proposition~\ref{prop: Strong equilibria}.

\begin{eg}
\label{example; strong equilibrium}
Consider a committee with three members operating under unanimity, $\k=3$ and $\h \k =1$, with $u=(-4,2,3)$.    Proposition~\ref{prop: Strong equilibria} implies that the transfer $t^0$ is not a strong equilibrium because $u_1 = -4 <0$ and $\h \k =1$. This occurs because members $2$ and $3$ can create a blocking coalition of the transfer $t^0$  by compensating member $1$ to support the reform, thereby enabling the committee to enact it. Consider now the one-sided  transfer $t$ defined by $t_{21}(R) = 1$, $t_{31}(R) = 3$ and zero otherwise. The associated net transfers are given by $(r^t =(4,-1,-3), s^t=0)$, the resulting payoff is $\h \pi(t) = (0,1,0)$ and hence, $O(t) = R$. Moreover, the transfer $t$ defined above constitutes a strong equilibrium, as it satisfies the inequalities \reff{eq: Strong Nash Characterisation}.

\end{eg}

In the remainder of the paper, since each strong equilibrium can be linked to a one-sided strong equilibrium transfer, we will focus without loss of generality on one-sided transfer promises contingent on passing the reform and express the constraints of strong equilibria in terms of the net transfers $r \in \cP$.  It is however important to observe that for any net transfer, $r \in \cP$ there exist multiple transfer promises $t \in \cA^I_R$ such that $r^t =r$. To see this, consider
the linear system $ r^t_i := \sum_{j} t_{ji}(R) - \sum_j t_{ij}(R)$ with $I$ equations and $I^2 - I$
unknown consisting of $t_{ij}(R)$. Given the large number of unknowns, the system admits multiple solutions.
Consider any solution $t_{ij}(R)$ of that system and observe that the one-sided transfer profile $t_{ij}(R) +C$ is also a solution of the system of equations for any constant $C>0$. Then, for $C$ sufficiently large, we have $t_{ij}(R)+C\ge 0$, for all $i$ and $j$, and thus $\{t_{ij}(R)+C\}_{i,j}\in \cA_R^I$ is a one-sided transfer promises that produce the net promise $r$. To limit this multiplicity, we then consider the set of strong equilibria that minimizes total promised transfers, resulting in the set of one-sided strong   equilibria that achieves this minimization. We formally define this set in the next section, demonstrate its existence, and highlight the general properties of strong minimal equilibria.

\vspace{-0.3cm}
\section{Strong equilibria with minimal transfer (SM)} \label{section: SM equilibria}

\vspace{-0.3cm}

We now define the set of {\it Strong Minimal} (SM) net transfers $r \in \cP$ denoted by $\pmb{\cS \cM}_R$. We first denote by $\pmb \cS_R$ the set of strong equilibrium net transfers, defined as the set of net promises that can be supported by a one-sided strong equilibrium transfer:
$$
\pmb \cS_R = \{ r \in \cP ~\mid ~ \exists t \in \cA^I_R \mbox{ such that } t \in \pmb \cS \mbox{ and } r^t =r\}.
$$
For any $r \in \cP$, the total transfers associated to $r$ is defined as
 \bea
\label{equ: Total payment}
\left.\ba{c}
\cT_{r} := \frac{1}{2} \sum_{ \dbI} |r_i| .
\ea\right.
\eea
A net promises profile $r \in \cP$ contingent on the reform is a Strong Minimal equilibrium or ``SM" if (i) it is a strong equilibrium, $ r \in \pmb \cS_R$, and (ii) it minimizes the total promises transfer among all strong equilibria,
 $$
 \left.\ba{c}
  \pmb{\cS \cM}_R := \{ r \in \pmb{\cS }_R ~|~ \cT_{r} = \inf_{r' \in \pmb{\cS }_R} \cT_{\bm{r'}} \}.
\ea\right.  $$

The following proposition shows that the Strong  Minimal equilibria exist.
\begin{prop}
\label{prop-cEexistence}[\textbf{Existence}]
The set of SM equilibria net transfers $\pmb{\cS \cM}_R $ is a non-empty convex and compact subset of $\cP$.
\end{prop}

Relative to the set $\pmb \cS_R$ of Strong equilibria, the set of SM equilibria  $\pmb{\cS \cM}_R $ adds the restriction that the equilibria are achieved in the ``cheapest possible way." The set $\pmb{\cS \cM}_R $ is a subset of the set  $\pmb{\cS}_R $ as the following example illustrates.

\begin{eg}
\label{example: SNM first illustration}
 Consider a committee consisting of $3$ members, ruled by the majority rule, $\k = \h \k =2$, and with utilities $u = (-4, 1, 5)$. Inequality \reff{eq: Strong Nash Characterisation} shows that the net promise transfer $r \in \cP$ is a strong equilibrium, that is, $r \in \pmb \cS_R$, if and only if:
\beaa
r_1 + r_2 \ge 3, \q r_1+r_3 \ge -1,\q r_2 + r_3 \ge -6,\q r_1+r_2+r_3 =0.
\eeaa
The set $\pmb \cS_R $ is depicted as the light blue triangle in Figure~\ref{Fig: Illustration}. Moreover, for any $r \in \pmb \cS_R$,
$
\cT_{r} =\frac{1}{2} (|r_1| + |r_1+r_3| + |r_3| ) \geq |r_3|  \geq 3,
$
and the inequality holds as an equality  
if and only if $r_3 = -3 $ and $2 \leq r_1 \leq 3$. Thus,  the set of  strong minimal equilibria $\pmb{\cS \cM}_R$, is represented by the horizontal side of the blue triangle of Figure~\ref{Fig: Illustration} bounded by the points $(2,-3)$ and $(3,-3)$, with $\cT_r = 3$.

\begin{figure}[htbp]
    \centering
\begin{minipage}{0.7\textwidth}
    \begin{tikzpicture}
        \begin{axis}[
            width=12cm,
            height=6cm,
            axis on top,
            xmin=1, xmax=6.5,
            ymin=-7, ymax=-2,
            xtick={1,2,3,4,5,6,7},
            ytick={-7,-6,-5,-4,-3,-2},
            xlabel={$r_1$},
            ylabel={$r_3$},
            axis x line=bottom,
            axis y line=left,
            xlabel style={anchor=north},
            ylabel style={anchor=east, rotate=-90},
            tick style={draw=none},
            major grid style={dashed, gray!50},
            legend style={
                at={(axis cs:2.1,-4.8)},
                anchor=south west,
                draw=none,
                fill opacity=0.8,
                text opacity=1,
                fill=white
            }
        ]

        \coordinate (highlightStart) at (axis cs:2,-3);
        \coordinate (highlightMid) at (axis cs:2.5,-3);
        \coordinate (highlightMid1) at (axis cs:4,-4);
        \coordinate (highlightEnd) at (axis cs:3,-3);

        \draw[-{Stealth[length=4mm, width=3mm]}, line width=1.5pt, color=black]
            (highlightMid) to[out=120, in=140, looseness=4.2]
            (axis cs:2.4,-4.5) node[
                font=\small,
                align=center,
                below=3pt
            ] {
                Strong minimal \\ equilibria $\pmb{\cS\cM_R} $
            };

        \addplot[fill=cyan!50, draw=none] coordinates {(2,-3) (6,-7) (6,-3)};

        \fill (2,-3) circle (3pt);
        \fill (3,-3) circle (3pt);
        \draw[line width=2pt] (2,-3) -- (3,-3);

        \fill[fill=cyan!50] (rel axis cs: 0.20, 0.020) rectangle (rel axis cs: 0.26, 0.075);
        \node[anchor=west, scale=0.8] at (rel axis cs: 0.26, 0.046) {Strong equilibria $\pmb{\cS_R}$};

        \draw[gray, dashed, thin] (2,-7) -- (2,-3);
        \draw[gray, dashed, thin] (3,-7) -- (3,-3);
        \end{axis}
    \end{tikzpicture}
\end{minipage}

\caption{\small  We consider the case $u = (-4, 1, 5)$ of a committee operating under majority rule covered in Example~\ref{example: SNM first illustration}. The figure describes the set of one sided equilibria ($\pmb{\cS }_R$ and $\pmb{\cS \cM}_R$) in the plane $(r_1,r_3)$.}
\label{Fig: Illustration}
\end{figure}

\end{eg}

Example~\ref{example: SNM first illustration} shows that multiple strong equilibria of net transfers persist even when focusing on those that minimize total transfers (\ref{equ: Total payment}). However, examining these strong minimal equilibria reveals general properties of the net transfers. The next proposition identifies a {\it critical voter}, a reform supporter $k_* \in \cC^R$, such that all promisers belong to the coalition $\{k_*, \cdots, I\}$ and all recipients to the coalition $\{1, \cdots, k_* - 1\}$.

\begin{prop}
\label{prop-order}[\textbf{Top-down equalizing transfers}] There exists a critical voter $k_*\in \cC^R$ such that any SM equilibrium net transfers profile  $r \in  \pmb{\cS \cM}_R \backslash \{ 0 \}$ satisfies
\bea
\label{order}
 -u_j \le r_j \le 0 \le r_i,\q  \mbox{and}\q u_i + r_i \le u_j +r_j \q\mbox{for all}\q i < k_* \le j,
\eea
and the total transfers associated to the SM equilibrium $ r$ satisfies:
\bea
\label{cTr+-}
\left.\ba{c}
\cT_{ r} = \sum_{i< k_*} r_i = \sum_{j\ge k_*} (-r_j).
\ea\right.
\eea
\end{prop}

Proposition~\ref{prop-order} shows that, in any SM equilibrium, only reform supporters with utility at least as high as that of the critical voter $k_*$ can promise transfers. These supporters are constrained by the individual rationality condition $-u_j \leq r_j$ for $j \geq k_*$, preventing transfers that exceed their utility gain. This condition is enforced by minimality. For illustration, in example~\ref{example: SNM first illustration}, the net transfer $r = (6, 1 - 7)$ forms a strong equilibrium with payoff $\hat \pi = (2,2,-2)$. Here, member 3 transfers $7$, exceeding her maximal utility of $u_3 = 5$. Such “excessive” transfers can arise because unilateral retractions are prohibited, but minimality rules them out.

Proposition~\ref{prop-order} shows that transfer recipients must belong to the coalition ${1, \cdots, k_* - 1}$, which may include both reform opponents and lower-utility reform supporters. In the latter case, higher-utility supporters transfer to lower-utility allies. The critical voter’s rank $k_*$ depends on the voting rule and utility distribution, and the SM equilibrium condition (middle inequality in \reff{order}) ensures that promisers in ${k_*, \cdots, I}$ maintain post-transfer utility no lower than any recipient in ${1, \cdots, k_* - 1}$. Thus, SM equilibria preserve utility rankings between promisers and recipients. Because $k_*$ is endogenous, Proposition~\ref{prop-order} remains agnostic about the size of transfers and the identities of participating members; total transfers depend on the utilities $u_i$ and the majority threshold $\k$. In the next section, we provide more  detailed descriptions of the SM equilibria.


\vspace{-0.3cm}

\section{Strong Minimal equilibria implications} \label{Section: Equilibrium promises}

\vspace{-0.3cm}

In this section, we examine both the magnitude and direction of SM equilibrium transfer flows and explore how multiplicity arises in various scenarios.  We first examine the case where the reform is defeated without transfers, and then discuss the scenario where the reform passes without transfers. The discussion and presentation of the principles are illustrated using specific numerical examples. However, these principles are general, and a detailed exposition of the general results, along with their proofs, can be found in the Supplemental Appendix SA. Before proceeding, we define the aggregate utility (resp. dis-utility) of the reform supporter (resp. opponents) 
by
\bea
\label{equ: aggregate intensity of preferences for R or for S}
U_R := \sum_{ \cC^R} | u_i | \equiv \sum_{ i = n+1}^I  u_i , \q U_S := \sum_{ \cC^S} | u_i | \equiv \sum_{i=1}^n (- u_i ).
\eea

\subsection{SM Equilibria when the reform is defeated without transfers} \label{section: frustrated minority}

\vspace{-0.2cm}

We examine scenarios where the reform lacks sufficient support without transfers, specifically when $\mid \cC^R \mid < \k$ and $O(t^0)=S$. In this context, the critical voter from Proposition~\ref{prop-order} is identified as $\k_*=n+1$. Thus, only the opponents of the reform can receive transfers. The minimum total transfer is given by $\cT_r= U_S$, indicating that reform supporters must compensate reform opponents sufficiently to cover their total utility loss from the adoption of the reform. Additionally, Proposition~\ref{prop-order} demonstrates that, following these transfers, the utility of any reform opponent cannot exceed that of any reform supporter. In general, there are multiple SM equilibria because there are various ways to allocate the cost of transfers among reform supporters and different methods for distributing the transfers among reform opponents. However, when reform supporters are at least two members short of reaching the quota $\k$, all equilibria involve compensating each reform opponent sufficiently to make each one of them indifferent between the reform and the status quo,  $r_i = -u_i$ for all $i \in \cC^S$. Finally, transfer recipients are always reform opponents and thus all transfers are reaching ``across the aisle". In cases of weak voting support for the reform, promises involving “circle the wagon” transfers where reform supporters compensate other reform supporters are, therefore, not expected to occur in SM equilibria. The analysis in full generality is done in Proposition~\ref{prop: reaching across the aisle} from the Supplemental Web Appendix $SB$  where we provide necessary and sufficient conditions under which all SM equilibrium promises are of the across the aisle type. To illustrate the subcase where the reform is defeated without transfers, consider an example.


\begin{eg}
\label{example: SNM}
Consider a committee consisting of three members, with utilities
$u = (-2, -1, 10)$. We have $\cC^R = \{ 3 \}$, $\cC^S = \{ 1,2 \}$, and $U_S =3$. In an SM equilibrium, member $3$ must transfer $3$ units of utility to members $1$ and $2$, regardless of whether the committee operates on a majority or unanimity basis.

When the committee operates under the unanimity rule, the coalition $\mathcal{C}^R$ is two members short of being decisive. In this case, all SM transfer promises $t$ yield the same net transfer $r = (2, 1, -3)$ and the same post-transfer utilities $\hat{\pi}(t) = (0, 0, 7)$.

When the committee operates under a majority rule: $\k = \h \k =2$, there are multiple SM equilibria because the coalition $\cC^R$ is one member short of reaching the quota $2$.  The transfer of $3$ units of utility initiated by member $3$ can be distributed among members $1$ and $2$ while adhering to the right inequality in (\ref{order}). For example both net transfers $r=(3,0,-3)$ and $r=(0,3,-3)$ are SM equilibria.

\end{eg}

\vspace{-0.3cm}
\subsection{SM equilibria when the reform is adopted without transfers}

\vspace{-0.3cm}

We now assume that the reform has sufficient support to be enacted without transfers, $O(t^0) = R$ or equivalently that the number of reform supporters is large, $\mid \cC^R \mid \geq \k$. 
 Members of the coalition $\cC^S$ can promise transfers contingent on defeating the reform to some members of the coalition $\cC^R$ to increase the support for their preferred policy $S$ and lead the committee to defeat the reform. The members of the coalition $\cC^R$ who are more susceptible to being enticed to vote for the status quo are those with the lowest utilities in the reform. When the enticements are persuasive, some reform supporters are converted to reform opponents, and these conversions represent a pivotal event if a blocking coalition exists. We denote the coalition of these reform supporters by
\be \label{eq: coalition of weakest reform supporters}
\underline \cC ^R := \{ n +1,..,\h\k\}.
\ee

Since $\cC^S \cup \underline \cC ^R = \{ 1,..,\h\k \}$, members of $\underline \cC^R$ are numerous enough to defeat the reform if they vote against it with the reform opponents. Denote by
$\underline U^R := \sum_{\underline \cC^R} u_i \equiv \sum_{i = n+1}^{\h \k} u_i$ the aggregate utility for the reform of the coalition $\underline \cC^R$.

For the members of the coalition $\cC^S$, the incentive to entice members of the coalition $\underline \cC^R$  to vote for the alternative $S$ arises only when there are gains from trade. Specifically, by switching from the decision $R$ to $S$, members of the coalition $\cC^S$ gain the aggregate amount $U^S$. To entice members of the coalition $\underline \cC^R$ to vote for $S$, members of the coalition $\cC^S$ incur the aggregate utility cost $\underline U^R $. Therefore, we define the aggregate gains from trade of the members of the coalition $\cC^S$ as
\be \label{eq: gain from trade}
G^S := U^S - \underline U^R = - \sum_{i\in \cC^S\cup \underline \cC^R} u_i \equiv -\sum_{i = 1}^{\h \k} u_i.
\ee
The sign of $G^S$ is the key factor in characterizing  the equilibrium transfer.

When the gain from trade $G^S$ is non-positive, the only SM equilibrium is the degenerate transfer $t^0$ since the reform opponents have no incentives to entice the members of the coalition $\underline \cC ^R $ to vote against the reform.
When $G^S >0$, transfers are necessary to prevent coalitions of reform opponents from blocking the transfer $t^0$, enticing some reform supporters into opposing the reform and coercing the remaining reform supporter.

Assuming $\mid \cC^R \mid \geq \k$ and $G^S>0$, two cases can occur. To describe them, 
we define the aggregate surplus of utility of the coalition $\cC^R \backslash \underline \cC^R$ relative to member  $\h k \in \underline \cC^R$ by
\bea
\label{Delta Ui}
\D U_{\h \k} := \sum_{j\in \cC^R \backslash \underline \cC^R} [u_j - u_{\h \k}]\equiv \sum_{j= \h \k+1}^{I} [u_j - u_{\h \k}].
\eea
The variable $\D U_{\h \k}$ represents the maximum aggregate transfer that members of the coalition $\cC^R \backslash \underline \cC^R$ can promise while keeping their after transfer utilities above that of member $\h \k$.

\vspace{-0.3cm}
\subsubsection{Case $1$: First order preemption.} \label{section: first order}

\vspace{-0.3cm}

We assume that  $\mid \cC^R \mid \geq \k$, $G^S>0$ and $\D U_{\h\k} \geq G^S$. When $\D U_{\h\k} \geq G^S$, the members of the coalition $\cC^R \backslash \underline \cC^R$ can afford to promise a total transfer of $G^S$ while maintaining their post-transfer utilities above that of all the members of the coalition of promises $\cC^S \cup \underline \cC ^R$. In this case, members of the coalition $\cC^S \cup \underline \cC ^R$ can be compensated for the forgone gain realized by forming a blocking coalition, i.e., $G^S$, without creating new targets for enticement against the reform. The critical voter is thus $k_* =\h \k +1$ and all equilibria involve a total transfer  of $\cT_r = G^S$ from the coalition $\cC^R \backslash \underline \cC^R =\{ \h \k+1,\cds,I\}$  to the coalition $\cC^S \cup \underline \cC ^R=\{ 1,\cds,\h \k\}$. In all SM equilibria,  the post-transfer individual utilities  remain larger for all members of $\cC^R \backslash \underline \cC^R$ than that of any member of the coalition $\cC^S \cup \underline \cC ^R$. To illustrate this case, we examine an example.

\begin{eg}
\label{ex: first order}
Consider a committee with $4$ members governed by the majority rule $\k = 3$ and utilities $ u = (-5,~1, ~2, ~ 10)$. In this case, $\h\k=2$, $\cC^S= \{1\}$, $\cC^R= \{2,~3,~4\}$,  $\underline \cC^R = \{2\}$, $\cC^R/\underline \cC^R = \{3,~4\}$, and $G^S = 4$.
 The aggregate surplus of utility of the coalition $\{3, 4\}$ is $\D U_{2} = (2-1)+(10-1)=10 > G^S =4$.  Therefore the coalition $\{3, 4\}$ can afford to promise a transfer of $4$ without violating the ordering condition of the post-transfer utilities from Proposition~\ref{prop-order}. In any SM  equilibrium, the coalition $\cC^R/\underline \cC^R= \{3, 4\}$ should promise a total transfer of  $G^S=4$ to the coalition $\{1, 2\}$. The critical voter from Proposition~\ref{prop-order} is thus $k_*= 3$. There are multiple SM equilibria.  For example, the net transfers $r = (3, ~1,~ 0,~ -4)$ is an SM equilibrium. It produces the post-transfer utilities $\h \pi(r) = (-2,~ 2,~ 2,~ 6)$ and includes a ``circle the wagon" transfer, as it involves a transfer from reform supporter $4$ to another reform supporter, member $2$, who has a weaker utility. The net transfer $r' = (4,~ 0, ~0, ~-4)$ is also an SM equilibrium as it produces the post-transfer utilities  $\h \pi(r')  = (-1,~ 1,~ 2,~ 6)$. The net transfer $r'' = (2,~ 2,-1,-3)$ achieves the minimum total transfer promises $\cT_{r''}=4$ but is not an SM equilibrium. This is because the post-transfer utilities $\h \pi(r'') = (-3, ~3,~ 1,~ 7)$
violate the ordering condition  $\h \pi_2(r'') \leq \h \pi_3(r'')$. Since $\h \pi_1(r'') + \h \pi_3(r'') =-2 <0$ after the net transfers $r''$ are implemented, member 3 becomes a new target for enticement by member 1.  Thus the coalition $\{1\}$ can block the net transfer $r''$, indicating that the net transfer $r''$ is not a strong equilibrium.
\end{eg}

\vspace{-0.5cm}
\subsubsection{Case 2: Higher order preemption} \label{section: second order}

\vspace{-0.2cm}

We now assume that  $\mid \cC^R \mid \geq \k$, $G^S>0$ but this time $\D U_{\h\k} < G^S$. When $\D U_{\h\k} < G^S$, if the members of the coalition $\cC^R \backslash \underline \cC^R$  promise a total transfer of $G^S$ to compensate the coalition $\cC^S \cup \underline \cC ^R$ for not undermining the reform, they violate the ranking of post-transfer utilities. Consequently, some members of the coalition $\cC^R \backslash \underline \cC^R$ become targets for new transfers aimed at enticing them to vote against the reform. These additional opportunities of transfers will require second-order compensation. For this reason, it is intuitive that the total transfer of SM equilibria must be larger than $G^S$. The results of Proposition \ref{prop: Equilibria in presence of gain from trade2 Alternative} from the supplemental Web Appendix SA show that the critical vote trader is a member $k_* \in \underline \cC^R$ defined by
\bea
\label{k*}
 u_{k_* -1}  \le u_* <u_{k_*},\q\mbox{where}\q u_* := {1\over \k-1}\sum_{j\in \dbI} u_j > 0,
\eea
and that the total transfer for any SM equilibrium net transfer $r$ is given by $\cT_{r} =\cT_*:=\sum_{j\ge k_*} \big[u_j - u_* \big] > G^S$. To illustrate this case, we discuss an example.


 \begin{eg}
\label{ex: second order}

Consider a slight variation of Example~\ref{ex: first order}, where a committee with $4$ members is ruled by majority rule, $\k = 3$, and where the utilities are now given by $ u = (-5,~1, ~2, ~ 3)$. In this case $\h\k=2$, $\cC^S= \{1\}$, $\cC^R= \{2,~3,~4\}$, $\underline \cC^R = \{2\}$, $\cC^R/\underline \cC^R = \{3,~4\}$, and $G^S = 4$. We have $\D U_{\h \k} \equiv \D U_{2}  = 3 < 4=G^S$ and thus the coalition $\{ 3,4\}$ cannot afford to promise the transfers of $4$ without creating new targets of enticement. In the first step, members $3$ and $4$  will promise a transfer to member $1$ in order to equalize their utilities with that of member $2$. The net transfers are therefore $r^1 = (3,~0,~-1,~-2)$, which result in the post transfer payoff $\h \pi (r^1) = (-2,~1, ~1, ~ 1)$. The net transfer is not an SM equilibrium because it is not strong Nash: $\h \pi_1 (r^1) + \h \pi_2 (r^1) = -1<0$, and thus additional transfer promises need to be made. Using the symmetry of this specific example, we conjecture that for an SM equilibrium, each member of the coalition $\{ 2,3,4 \}$ must transfer an amount $x$ to member $1$, which results in the net transfer $r^2=(3x,~-x,~-x,~-x)$ and thus the post-transfer utilities $\h \pi(r^1+r^2) = (-2+3x,~1-x,~1-x,~1-x)$. If we select $x=1$, the net transfer $r^1+r^2 = (6,~-1,~-2,~-3)$ is a strong equilibrium in which total transfer $\cT_{r^1+r^2} = 6$ is too large. To achieve the minimum total transfer, we solve the equation $(-2+3x)+(1-x) =0$ and get the solution $x=1/2$. The net transfers promise $r^1+r^2 = (9/2,~-1/2,~-3/2,~-5/2)$ is an SM equilibrium with total transfer $\cT_{r^1+r^2} = 9/2 >4$ and payoff $\hat \pi (r^1+r^2)= (-\frac{1}{2},~\frac{1}{2},~\frac{1}{2},~\frac{1}{2})$. The critical voter is thus $k_*=2$.
\end{eg}

\vspace{-0.3cm}
\section{Discussion of dynamic implementations} \label{Section: Discussion}

\vspace{-0.3cm}

In our model, we assume that transfer promises are made simultaneously and do not explicitly model the game leading to equilibria. However, once an equilibrium transfer
profile is reached, it cannot be overturned, while other transfer profiles do not exhibit the same level of stability.
To understand how the equilibria are reached, it would be beneficial to model a dynamic game of sequential, decentralized transfer promises before voting. A strong equilibrium $t \in {\pmb \cS}$ can be viewed as the cumulative result of these sequential promises.

We consider a dynamic process of coalition formation, where sequences of transfers provide myopic, strict gains to all coalition members initiating them at each step. We define a Pivot Algorithm as a mechanism that generates a sequence of blocking coalitions as follows: Start with the degenerate transfer $t^0$. If no blocking coalition exists, the process stops. If blocking coalitions are present, one is chosen according to a selection rule. The transfer associated with the selected blocking coalition is denoted by $t^1$. We then consider a new transfer game where the status quo payoff is updated from $0$ to $(s_i^{t^1})_{i \in \mathbb{I}}$, and the reform payoff changes from $(u_i)_{i \in \mathbb{I}}$ to $(u_i + r_i^{t^1})_{i \in \mathbb{I}}$. In this game, post-transfer payoffs at each step become the starting payoffs of a new game. If there is no blocking coalitions in the second step, then stop. If blocking coalitions exist, then select one of them according to the given selection rule. The iteration continues until no further blocking coalitions remain, assuming convergence occurs. A Pivot Algorithm is thus defined by the selection rule used when multiple blocking coalitions exist. Since a blocking coalition must improve the payoff of its members at each step, the committee decisions are overturned in each step, making some transfer recipients pivotal—hence the term ``Pivot Algorithm". Additionally, as the game’s payoffs are updated in each step, the assumption of non-negative transfers implies that no retraction take place in the pivot algorithm.




The following proposition shows that any strong equilibrium net transfer is reachable through a Pivot-algorithm  under certain configurations of the payoff vector $u$.

\begin{prop}
\label{prop-dynamic}
Assume $t^0\notin \pmb\cS$.  Any net transfer $r\in \pmb{\cS \cM}_R$ is reachable through a Pivot algorithm in at most two steps under the following conditions:

(i) When $|\cC^R|<\k$, the net transfer $r$ is reachable in one step: $t^0$ is blocked by a coalition $\cC \subseteq  \cC^R$ with corresponding $t\in \pmb \cS$ satisfying $t(S) = 0$, $r^t = r$ and, $O(t)=R$.

(ii) When $|\cC^R| \ge \k$ and $\Delta U_{\hat \k} \ge G^S > 0$, the net transfer $r$ is reachable in two steps: In the first step, there exists a blocking coalition $\cC \subseteq \cC^S$ for the transfer $t^0$ with corresponding $t \in \cA^I$ that satisfies $O(t)=S$. In the second step, there exists a blocking coalition $\tilde \cC \subseteq \cC^R$ for the transfer $t$ with corresponding $\tilde t \in \cA^I$, such that $r^{\tilde t} = r$, $s^{\tilde t} = 0$ and, $O(\tilde t)=R$.
\end{prop}

Proposition \ref{prop-dynamic} states that SM equilibria can be implemented using a Pivot algorithm, where a suitable selection rule for blocking coalitions ensures convergence within at most two steps. However, different selection rules may lead to alternative equilibria or even cycles with diminishing transfers and non-convergence. The proposition establishes that certain selection rules can implement our equilibria through a Pivot algorithm but leaves open the question of what motivates the selection rule. Addressing this requires defining a coalition formation game  (\cite{ray2015coalition}). Alternatively, the selection rule can be viewed as an endogenous agenda-setting mechanism (\cite{dutta2004equilibrium}). Developing these aspects would require a significant extension of the model, which we leave for future research.
In the following example, we illustrate the Pivot algorithm selection in the case $(ii)$ of Proposition~\ref{prop-dynamic}.

\begin{eg}
\label{ex: Dynamic implementation}
Reconsider the committee with $4$ members in Example~\ref{ex: first order}, where decisions follow the majority rule $\k = 3$, and members have the utilities $u = (-5,~1,~2,~10)$. Without transfer, the reform is adopted $O(t^0) = R$.
Consider the net transfer $r=(3,~1,~0,~-4) \in \pmb{\cS \cM}_R$, which generates the payoff  $\h \pi(t) =(-2,~2,~2,~6 )$ for any $t$ satisfying $r^t = r$. We proceed now to implement the net transfer $r$ in two steps.

\no{\bf Step 1.} Member $1$ blocks the degenerate transfer $t^0$ with the deviation $t_{12}(S) = 1+\e$ where $\e$ is a small number. The resulting net transfers are $r^{t}=0$ and $s^{t} = (-1 - \e,~1+\e,~0,~0)$.  Since $1+\e >1$, the committee decision is overturned, $O(t)=S$, and the payoffs profile is $\h \pi(t) = (-1-\e,~1+\e,~0,~0)$. Since the utility of member $1$ improves from $-5$ to $-1-\e$, member $1$  forms the blocking coalition of the first step.

\no{\bf Step 2.}  Members $\{2, 4\}$ form a blocking coalition and promise the two-sided transfers $t'$ defined by
$$
t'_{21}(S)=1+\e,~t'_{41}(R)=3,~t'_{42}(R)=1, \mbox{ and zero otherwise}.
$$
Denote $\tilde t := t + t'$ and observe that the coalition $\tilde \cC = \{2, 4\}$ satisfies the condition  \reff{coalition-retraction} with the deviation $\tilde t$.
Notice that $s^{\tilde t} = 0$ because member 2 deviates by returning the transfer promised by member 1 in the first step. This step is necessary since we aim to implement a one-sided transfer promise where all promises are contingent on reform adoption. The net transfer contingent on the reform is $r^{\tilde t} = (3,~1,~0,~-4) \equiv r$. Hence $\tilde  t$ implements the SM equilibrium $r$, and $\h \pi(\tilde t) = (-2,~2,~2,~6 ) $. As a last step, we need to verify that the coalition $\{2,4\}$ improves the payoff of its members after deviating. This holds true because member $2$'s payoff increases from $1+\epsilon$ to $2$, while member $4$'s payoff rises from $0$ to $6$ and, hence the deviation $\tilde t$ results in a strict improvement of all coalition members' utility. As a result, the coalition $\{ 2,3\} $ blocks the transfer $t$ with the deviation $\tilde t$. Thus we have implemented the equilibrium net transfer $r$.
\qed
\end{eg}

\vspace{-0.3cm}

\section{Conclusion} \label{Section: Conclusion}

\vspace{-0.3cm}

In this paper, we demonstrate the feasibility of analyzing pre-vote transfer promises contingent on vote outcome within committee settings governed by a quota rule. We assume members can form coalitions, commit to outcome-contingent transfers, and follow cooperative norms. Under these conditions, strong Nash equilibria that minimize total transfers exist, making such promises efficiency-enhancing and mitigating majority coercion. Importantly, we provide a detailed characterization of equilibrium transfers. Understanding these transfers is key to designing institutions that support such promises and reduce associated transaction costs.

 Our results rely on the assumption of commitment and coordination. Mechanisms that support these features are therefore crucial to address the political failure of majority coercion through transfer promises. Recently, the rise of online democratic platforms has introduced new approaches to governance. A notable example is the Decentralized Autonomous Organization (DAO), where rules are encoded as smart contracts on a blockchain (see, e.g., \cite{hall2024happens} for a non-technical overview of DAOs). A blockchain is a technology designed to maintain shared records without relying on a central authority, while a smart contract is a self-executing agreement whose terms are directly written into code, allowing for automated and enforceable governance rules (see \cite{roughgarden2024computer}). This framework enables decentralized decision-making, with stakeholders collectively voting on issues such as fund allocation and project selection. Our analysis is broadly relevant in this context, as technology can facilitate the implementation of coordination and commitment mechanisms.

It is equally important to recognize that other forms of political failure—such as the tragedy of the commons (\cite{olson1971logic}), voter ignorance (\cite{downs1957economic}), and rent-seeking (\cite{tullock1967welfare})—remain significant and merit further investigation. Much work remains to be done to develop a comprehensive evaluation of transfer promises, particularly in environments where informational frictions play a central role.  

\vspace{-0.5cm}

\section{Appendix: Proofs of the paper's propositions}


\vspace{-0.3cm}

\no {\bf Proof of Lemma~\ref{lem: BC overturn}:}
 Consider a blocking coalition $\cC$ for the transfer $t$ that deviates with the  transfers $\tilde t$ so that $\h \pi_i(  \tilde t) > \h \pi_i(t)$ for all $i \in\cC$. Assume further that the committee decision is not changed by the coalition, $O(t) = O(\tilde t) = R$, for instance.  In that case, $\h \pi_i(t) = u_i +r^t_i$ and $\h \pi_i( \tilde t) = u_i +  r^{\tilde t}_i$ for all $i\in \cC$. Then $ r^{\tilde t}_i >  r^{t}_i$ for all $i \in \cC$.  On the other hand, for any $i\in \dbI/\cC$, equation \reff{coalition-retraction} implies $\tilde t_i = t_i$ and $\tilde t_{ji}(O) \ge t_{ji}(O)$ for all $j$ and $O$, and thus
 \beaa
  r^{\tilde t}_i = \sum_{j \in \mathbb{I}} \tilde t_{ji}(R) -\sum_{j \in \mathbb{I} } \tilde t_{ij}(R)  \ge \sum_{j \in \mathbb{I}}  t_{ji}(R) -\sum_{j \in \mathbb{I} }  t_{ij}(R) = r^t_i.
  \eeaa
  Then, since $\cC$ is nonempty,
$  \sum_{i\in \dbI} (r^{\tilde t}_i - r^t_i) = \sum_{i\in \cC} (r^{\tilde t}_i - r^t_i) + \sum_{i\in \dbI/\cC} (r^{\tilde t}_i - r^t_i) >0.$
  This contradicts with the zero sum conditions for the total net transfers  $ \sum_{i\in \dbI} r^{\tilde t}_i = \sum_{i\in \dbI} r^t_i = 0$.
A similar reasoning applies to the case where $O(t) = O(\tilde t) = S$ and we omit it here for brevity. \qed


\no {\bf Proof of Proposition~\ref{prop: Non existence of Nash}:} For any transfer profile $t \in \cA^3$, we order the vector $u^t$ as
$
u^t_{i_1} \geq u^t_{i_2} \geq u^t_{i_3}.
$ Noticing that $\sum_j u^t_{i_j} = 1$, we must have $u^t_{i_1} > 0$. In the sequel, we will consider all possible configurations of the vote outcome and show that there are always unilateral winning deviations, and therefore, no Nash equilibrium exists.

\no{\bf Case 1.} Assume that $O(t) = S$. Because the reform is defeated, it must be that member $2$ and $3$ vote against it (recall that the committee is ruled by the majority rule).  Then we have $u^t_{i_1}>0 > u^t_{i_2}\ge u^t_{i_3}$ which implies that  Member $i_1$ has a benefitting deviation: instead of using the transfer $t_{i_1,i_2}(R)$ and $t_{i_1,i_3}(R)$ , member $i_1$ can deviate with the transfer
$$
\tilde t_{i_1,i_2}(R) = t_{i_1,i_2}(R) + \frac{1}{3} - u^t_{i_2}>0 ,~~\tilde t_{i_1,i_3}(R) = t_{i_1,i_3}(R) + \frac{1}{3} - u^t_{i_3} >0
$$
and keep all other transfers contingent on $O=S$ unchanged. Notice that there is no retraction in this deviation since $\frac{1}{3} - u^t_{i_j} > 0$ for $j = 2, 3$. It can be checked that all committee members vote for $R$ after the deviations and that member $i_1$ receives the utility $s^t_{i_1} + \frac{1}{3}$ instead of $s^t_{i_1}$ prior to the deviation, implying that the transfer profile $t$ cannot be a Nash equilibrium.




\ms

\no{\bf Case 2.} $O(t) = R$ and $t_{ij}(R)>0$ for some $i \neq j$.

\ms

{\bf Case 2.1.} When $u^t_{i_3} \geq 0$, the reform receives unanimous support. Thus, $t \notin \mathcal{N}$, as member $i$ can benefit from retracting the promise $t_{ij}(R)$ since no member is pivotal.

{\bf Case 2.2.} When $u^t_{i_3} < 0 = u^t_{i_2}$, member $i_3$ can benefit from adding a small transfer $0 < \varepsilon < |u^t_{i_3}|$ to member $i_2$ contingent on $S$. Hence $t \notin \mathcal{N}$.


{\bf Case 2.3.} Assume $u^t_{i_3} < 0 < u^t_{i_2}$. If $j = i_3$, member $i$ benefits from retracting $t_{ij}(R)$ without affecting the vote, so $t \notin \mathcal{N}$.

Now, if $j = i_1$ or $i_2$ (say $j = i_2$), $t \notin \mathcal{N}$ as member $i$ benefits from partially retracting $t_{ij}(R)$. Defining $\varepsilon < \min(t_{ij}(R), u^t_{i_2})$ and setting $\tilde t_{ij}(R) = t_{ij}(R) - \varepsilon$ and $\tilde t=t$ otherwise, keeps $i_2$ voting for $R$, ensuring $O(\tilde t) = R$ and benefiting member $i$.

\ms
\no{\bf Case 3.} $O(t) = R$ and $t(R)\equiv 0$, and thus $r^t=0$ and  $u + r^t = u = (-3, 2, 2)$. In this case, $t \notin \mathcal{N}$ since member $1$ can benefit by changing her promises contingent on $S$. To describe the deviation, assume without loss of generality that $t_{23}(S) \le t_{32}(S)$. Consider the deviation $ \tilde t$ defined by
$
\tilde t_{12}(S) = t_{21}(S) + 2.5,\q \tilde t_{13}(S) = t_{31}(S)
$ and $\tilde t=t$ otherwise.
Then
\beaa
&&\dis  s^{\tilde t}_1 = t_{21}(S) + t_{31}(S) - \tilde t_{12}(S) - \tilde t_{13}(S) = - 2.5 > -3 = u_1 +  r^{\tilde t}_1;\\
&&\dis s^{\tilde t}_2 = \tilde t_{12}(S) + t_{32}(S) - t_{21}(S) - t_{23}(S) = 2.5+ t_{32}(S) - t_{23}(S) \ge 2.5 > 2 = u_2 +  r^{\tilde t}_2.
\eeaa
Then $O(\tilde t) = S$, and member $1$ get the benefit $3-2.5=0.5$ from the deviation.\qed



\no{\bf Proof of Proposition~\ref{prop: Strong equilibria}:} We proceed in four steps.

\no{\bf Step 1.} In this step we show that $\pmb\cS$ is nonempty. First, define $r\in \cP$ by:
$$
r_i = -u_i >0,\q i\le n;\q r_i := {\sum_{j\le n} u_j \over \sum_{j>n} u_j} u_i \le 0,~ i>n.
$$
 By viewing $\{t_{ij}(R)\}_{i\neq j}$ in equation \reff{equ: Net transfer} as unknowns and $r$ as given coefficients, it can be  shown by solving linear equations
that there exists $t\in \cA^I_R$ such that $r^t = r$. Note that $s^t \equiv 0$, and that $u_i^t:= u_i +r_i$ for any $i$, we have
$$
u^t_i =  0,~ i\le n;\q u^t_i =  \big[1+ {\sum_{j\le n} u_j \over \sum_{j>n} u_j}\big] u_i =  {\sum_{j\in \dbI} u_j \over \sum_{j>n} u_j} u_i \ge 0,~ i>n.
$$
Then, given $t$, all members in $\dbI$ vote for $R$.

We now prove that $t \in \pmb \cS$. Assume by contradiction that $t\notin \pmb \cS$, then there exists a blocking coalition with retractions $\cC$ with corresponding $\tilde t$, as defined in \reff{coalition-retraction}. Since $O(t) = R$, by Lemma \ref{lem: BC overturn} the blocking coalition $\cC$ must defeat the reform and thus $O(\tilde t) = S$. This implies $s^{\tilde t}_i = \hat \pi_i(\tilde t) > \hat \pi_i(t) = u^t_i \ge 0$ for all $i\in \cC$.
On the other hand, for $i\notin \cC$, we must have $\tilde t_i = t_i$ and $\tilde t_{ji}(S) \ge t_{ji}(S)$ for all $j$. Then
$s^{\tilde t}_i = \sum_j [\tilde t_{ji}(S) - \tilde t_{ij}(S)] \ge \sum_j [ t_{ji}(S) -  t_{ij}(S)]  = s^t_i =0.$
Put together, we have $\sum_{i\in \dbI} s^{\tilde t}_i = \sum_{i\in \cC} s^{\tilde t}_i +\sum_{i\notin \cC} s^{\tilde t}_i > 0.$
This contradicts with the fact that $ \sum_{i\in \dbI} s^{\tilde t}_i=0$ for all $\tilde t\in \cA^I$. Thus $t \in \pmb \cS$.

\no{\bf Step 2.} In this step, we show that $O(t) = R$ for all $t\in  \pmb \cS$. Assume by contradiction that $O(t) = S$ for some $t\in  \pmb \cS$. Introduce
$
r'_i :=  \ol u_* - u^t_i,$  where $ \ol u_* := \frac{1}{I} \sum_{j\in \mathbb{I}} u_j >0
$ and where we recall that $u_i^t = u_i +r^t_i - s^t_i$ is the post-transfer net utilities.
Note that $\sum_{i\in \dbI} r'_i=0$ and hence there exists $ t'\in \cA^I_R$ such that $r^{t'} = r'$. Moreover, denoting $\tilde t := t+t'$, then
\beaa
u^{\tilde t}_i = u^t_i + r^{t'}_i - s^{t'}_i = u^t_i + r'_i = \ol u_* > 0,\q i\in \dbI.
\eeaa
Therefore, $O(\tilde{t}) = R$, and since $O(t) = S$, we have:
\beaa
\hat\pi_i(\tilde t) - \hat\pi_i(t) = (u_i + r^{\tilde t}_i) - s^t_i = u^t_i +  r'_i = \ol u_* > 0,\q  i\in \dbI.
\eeaa
In particular, the above holds true for $i\in \cC$ for the $\cC$ defined by \reff{coalition-retraction}. That is, $\cC$ is a blocking coalition for $t$, a desired contradiction with $t\in \pmb\cS$. Therefore, $O(t) = R$.

\ms
\no{\bf Step 3.} In this step, we prove that if $t \in \pmb{\cS}$, then the inequalities on the right-hand side of (\ref{eq: Strong Nash Characterisation}) hold. We proceed by contradiction and assume that there exists a coalition $\tilde\cC$ satisfying $|\tilde \cC | \geq \h \k$   such that $\sum_{i\in \tilde\cC} u^t_i < 0$. Following similar arguments from Step 2, by restricting to all members in $\tilde\cC$ and reversing the roles of $R$ and $S$, there exist $t'\in \cA^I$ such that $t'_{ij}=0$ if either $i$ or $j$ is not in $\tilde\cC$, $ t'(R) \equiv 0$, and $u^t_i + r^{t'}_i = u^t_i < s^{t'}_i$ for all $i\in \tilde\cC$. Denote $\tilde t := t + t'$. Since $|\tilde\cC|\ge \hat\k$, this implies that $O(\tilde t) = S$. From Step $2$, we know $O(t)=R$ and hence,
$
\hat\pi_i(\tilde t) - \hat\pi_i(t) = s^{\tilde t}_i - (u_i + r^t_i) = s^{t'}_i - u^t_i >0$, for all $ i\in \tilde\cC.
$
Moreover, let $\cC$ be the coalition defined by \reff{coalition-retraction}. Since $t'_{ij}=0$ if either $i$ or $j$ is not in $\tilde\cC$, then we must have $ \cC\subset \tilde\cC$, and thus $\hat\pi_i(\tilde t) - \hat\pi_i(t)  >0$ for all $ i\in \cC$. This implies that $\cC$ is a blocking coalition for $t$, contradicting with the assumption that $t\in \pmb\cS$.

\no {\bf Step 4.} We prove that if a promises profile $t\in \cA^I$ satisfies the inequalities (\ref{eq: Strong Nash Characterisation}) for every coalition of size $|\cC| \geq \h \k$, then  $t\in \pmb \cS$.
 First, if inequalities (\ref{eq: Strong Nash Characterisation}) hold, then $O(t) = R$. Indeed, if $O(t) = S$, a coalition $|\cC| \geq \h \k$ must unanimously support the status quo. However, this is impossible since $\sum_\cC (u_i + r^t_i) \ge \sum_\cC s^t_i$ contradicts $u_i + r^t_i < s^t_i$ for all $i \in \cC$.


To prove $t \in \pmb \cS$, we argue by contradiction. Assume there exists a blocking coalition  with retractions $\cC$ with corresponding $\tilde t$, as defined in \reff{coalition-retraction}. Since $O(t) = R$, any blocking coalition must overturn the reform, meaning $O(\tilde{t}) = S$ and thus $|\tilde{\cC}| \geq \hat{\k}$, where $\tilde{\cC} := \{i: u^{\tilde{t}}_i < 0\}$.
 Noticing that $|\cC \cup \tilde \cC| \ge |\tilde \cC| \ge \hat\k$, the inequality (\ref{eq: Strong Nash Characterisation}) applied to $\cC \cup \tilde \cC$ gives
\bea
0 \le \sum_{i\in \cC \cup \tilde \cC} u^t_i  = \sum_{i\in \cC} u^t_i +\sum_{i\in \tilde \cC\backslash \cC} u^t_i.
\label{contradiction Appendix}
\eea
For $i\in \cC$, by the definition of blocking coalition, we have $\hat\pi_i(\tilde t) = s^{\tilde t}_i > u_i + r^t_i = \hat \pi_i(t)$, then $u_i^t = u_i + r^t_i - s^t_i < s^{\tilde t}_i - s^t_i$.  For $i \in \tilde \cC\backslash \cC \subset \tilde \cC$, since $u^{\tilde{t}}_i < 0$, we have $u^t_i < u^t_i - u^{\tilde t}_i = (s^{\tilde t}_i - s^t_i) - (r^{\tilde t}_i - r^t_i)$. Then by \reff{contradiction Appendix} we have
\bea
0 &<& \sum_{i\in \cC} ( s^{\tilde t}_i - s^t_i ) +\sum_{i\in \tilde \cC\backslash \cC} \big[ (s^{\tilde t}_i - s^t_i) - (r^{\tilde t}_i - r^t_i)\big]\nonumber\\
&=&  \sum_{i\in \cC \cup \tilde \cC} (s^{\tilde t}_i - s^t_i) - \sum_{i\in \tilde \cC\backslash \cC}  (r^{\tilde t}_i - r^t_i) = - \sum_{i\notin \cC \cup \tilde \cC} (s^{\tilde t}_i - s^t_i) - \sum_{i\in \tilde \cC\backslash \cC}  (r^{\tilde t}_i - r^t_i),
\label{contradiction Appendix2}
\eea
where the last inequality is due to the fact that $s^t, s^{\tilde t}\in \cP$. For each $i\notin \cC$, by \reff{coalition-retraction} we see that  $\tilde t_{ij}(O) = t_{ij}(O)$ and $\tilde t_{ji}(O) \ge t_{ji}(O)$ for all $j\in \dbI$ and $O=R, S$, and thus $s^{\tilde t}_i \ge s^t_i$ and $r^{\tilde t}_i \ge r^t_i$. Then the last term of \reff{contradiction Appendix2} must be non-positive, which contradicts with \reff{contradiction Appendix2}.
\qed

\no {\bf Proof of Proposition~\ref{prop-cEexistence}:}  First, note that $\cS_R = \{r^t-s^t: t\in \pmb\cS\}$. Then by Proposition~\ref{prop: Strong equilibria} we see that $\pmb \cS_R$ is non-empty and closed, with
\bea
\label{eq: Strong Nash Characterisation2}
\left.\ba{c}
\pmb \cS_R = \big\{r\in \cP: \sum_{i\in \cC} (u_i + r_i) \ge 0 ~\mbox{for all coalitions}~ |\cC|\ge \hat\k\big\}.
\ea\right.
\eea
Fix an arbitrary $r^0\in \pmb \cS_R$, and denote $\pmb \cS_R(r^0) := \{r\in \pmb \cS_R: \cT_r \le \cT_{r^0}\}$.  Since $\cT_r > \cT_{r^0}$ for $r\in \pmb \cS_R \backslash \pmb \cS_R(r^0)$, then $ \pmb{\cS \cM}_R = \{r\in  \pmb \cS_R(r_0): \cT_r = \inf_{r'\in \pmb \cS_R(r^0)} \cT_{r'}\}$. The continuity of the function $r\to \cT_r$ implies that $\pmb \cS_R(r^0)$ is closed and bounded, and thus compact.  Consequently, the set of minimizers $\pmb{\cS \cM}_R$ is non-empty and compact. Finally, for any $r^1, r^2 \in   \pmb{\cS \cM}_R \subset \pmb{\cS}_R$ and any $0<\a<1$, denote the convex combination $r:= \a r^1 +(1-\a) r^2 \in \cP$. Note that if $r^1, r^2$ satisfy  \reff{eq: Strong Nash Characterisation2},  $r$ also satisfies \reff{eq: Strong Nash Characterisation2} and hence $r\in \pmb\cS_R$. Moreover, since $r^1$, $r^2$ are minimizers, we have $\cT_r \le \a \cT_{r^1} + (1-\a) \cT_{r^2}= \inf_{r'\in \pmb \cS_R} \cT_{r'}$. Thus $r\in \pmb{\cS \cM}_R$. This proves that $\pmb{\cS \cM}_R$ is convex.
\qed


\no{\bf Proof of  Proposition~\ref{prop-order}:}  For convenience, we use the notation $O(r) = O(t)$ and $\h \pi (r) = \h \pi (t)$, assuming $r^t = r$. The proof relies on the   following three lemmas.

\begin{lem}
\label{lem-switchorder}
For any equilibrium net transfer promises $ r \in \pmb{\cS \cM}_R \backslash \{0\}$, there exists no pair of members $(i, j)$ such that
\bea
\label{switch}
r_i < 0 < r_j \q\mbox{and}\q \h \pi_i(r) < \h \pi_j(r).
\eea
\end{lem}

\no{\bf Proof :} Assume by contradiction that \reff{switch} holds true. For some small $\e>0$, set
\bea
\label{tilder}
\tilde r_i = r_i + \e\le 0, \q \tilde r_j = r_j -\e\ge 0,\q\mbox{and} \q\tilde r_k = r_k~ \mbox{for all}~ k\neq i, j.
\eea
Notice $\sum_\dbI \tilde r_i = \sum_\dbI  r_i =0$ and hence, $\tilde r\in \cP$. Next, for any $|\cC| \geq \h \k$,  since $ r \in \pmb{\cS \cM}_R  \subset \pmb{\cS}_R$,   using  the characterization \reff{eq: Strong Nash Characterisation2}
and  inequality \reff{switch}  gives
\bea
\label{tilderstable1}
\sum_{k\in \cC} ( u_k + \tilde r_k )  = \left\{\ba{lll}
\dis \sum_{k\in \cC} ( u_k +  r_k ) \ge 0,\q\mbox{if}~ i, j \in \cC~\mbox{or}~ i, j\notin \cC;\ms\\
\dis   \sum_{k\in \cC}  ( u_k +  r_k )  + \e > \sum_{k\in \cC}  ( u_k +  r_k ) \ge 0,\q \mbox{if}~ i \in \cC, j\notin \cC.
\ea\right.
\eea
For the last case that $j \in \cC, i\notin \cC$,  by setting $ \e < \h \pi_j(r) - \h \pi_i(r)$, we have $\h \pi_i(r) < \h \pi_j(r)  - \e =  u_j + \tilde r_j$. Then, since the cardinal of the coalition $(\cC\backslash \{j\}) \cup\{i\}$ is also larger than $\h \k$, we have
\bea
\label{tilderstable2}
\sum_{k\in \cC} ( u_k + \tilde r_k )   \ge \sum_{k\in (\cC\backslash \{j\}) \cup\{i\}} ( u_k + \tilde r_k ) \ge 0, \q \mbox{if}~ j \in \cC, i\notin \cC.
\eea
Combining \reff{tilderstable1} and \reff{tilderstable2} and recalling that $\cC$ is an arbitrary coalition satisfying $|\cC| \geq \h \k$,  by \reff{eq: Strong Nash Characterisation2} again
we see that $\tilde r \in \pmb \cS_R$. Note further that
\beaa
|\tilde r_i| = - \tilde r_i = - r_i -\e= |r_i| - \e,\q  |\tilde r_j| =  \tilde r_j = r_j -\e= |r_j| - \e,\q |\tilde r_k| =  |r_k|, ~k\neq i, j.
\eeaa
 Then we obtain the following desired contradiction with the minimum  property of $ r\in \pmb{ \cS \cM}_R$:
\beaa
\cT_{\tilde r} = {1\over 2}\sum_{k=1}^I |\tilde r_{k}| = {1\over 2} \Big[\sum_{k\neq i, j} |r_{k}| + |r_{i}| - \e +|r_{j}| - \e\Big] = \cT_{ r } - \e <  \cT_{ r }.
\eeaa

\vspace{-1.2cm}
\qed

\bs
\begin{lem}
\label{lem: intermediary step}
For any 
$r \in \pmb{\cS \cM}_R \backslash \{ 0\}$, there exists no committee member $i$ such that
\bea \label{eq: sufficient condition for contradiction}
 r_i<0 \q\mbox{ and } \q \h \pi_i(r)  <0.
\eea
\end{lem}

\no{\bf Proof :} Assume by contradiction that \reff{eq: sufficient condition for contradiction} holds true. Since $ r\in \cP$, there exists $j\neq i$ such that $r_j>0$. Then by Lemma \ref{lem-switchorder}, we have $\h \pi_j(r)  \le \h \pi_i(r)   <0$.  Define the net transfer  $\tilde r$ by  \reff{tilder} again. Note that $u_i + \tilde r_i = \h \pi_i(r)  +\e$,  $u_j + \tilde r_j = \h \pi_j(r)  - \e$. Then for $\e>0$ small enough,  we have $u_j+ \tilde r_j < u_i + \tilde r_i  \le 0$. Following the same reasoning as in Lemma \ref{lem-switchorder}, we prove now that the inequality \newline $\sum_{k \in \cC} (u_k+ \tilde r_k) \geq 0$ must hold for every coalition $|\cC| \geq \h \k$.

First, observe that   by \reff{eq: Strong Nash Characterisation2},
we have  $\sum_{k\in \cC} \h \pi_k(r)  \ge 0$ and hence,  there exists $m \in \cC$ such that $\h \pi_m(r)  \ge 0$. Notice that, since $\h \pi_j(r)  \le \h \pi_i(r)  <0$, we have $m\neq i, j$.
When $i, j \in \cC$, or $i, j \notin \cC$, or $i\in \cC, j\notin\cC$, \reff{tilderstable1} remains true and hence $\sum_{k \in \cC} (u_k+ \tilde r_k) \geq 0$.
 In the last case where $j\in \cC, i\notin \cC$, since  $u_m + \tilde r_m = u_m + r_m \ge 0 \ge u_i + \tilde r_i$ and $|\cC\backslash \{m\}) \cup\{i\} | \geq \h \k$, we have
\beaa
\sum_{k\in \cC} ( u_k + \tilde r_k )    \ge \sum_{k\in (\cC\backslash \{m\}) \cup\{i\}} ( u_k + \tilde r_k )  = \sum_{k\in (\cC\backslash \{m\}) \cup\{i\}} ( u_k + r_k )\ge 0.
\eeaa
This, together with \reff{tilderstable1}   and \reff{eq: Strong Nash Characterisation2},
implies $ \tilde r \in \pmb \cS_R$. Then, following the same argument on the total transfer used in the last step of the proof of Lemma \ref{lem-switchorder}, we derive the desired contradiction.
\qed

\begin{lem}
\label{lem-overline k underline k} For any equilibrium promises profile  $ r\in \pmb \cS \backslash \{ 0 \}$,  introduce 
\bea
\label{ulolkappa}
\ul k_*  := \max \{ i | r_i > 0 \},\qq \ol k_*  := \min \{ i | r_i<0 \}.
\eea
 Then $\ul k_* < \ol k_*$, and
\bea
\label{ulolkappa2}
r_i\ge 0,~\mbox{for all } ~i \le  \ul k_*;\q r_k = 0,~\mbox{for all}~  \ul k^*< k < \ol k_*;\q  -u_j \le r_j \le 0,~ \mbox{for all }~ j \ge \ol k_*.
\eea
That is,  none of the committee members $i \le \ul k_*$ is a net  promiser; committee member $\ol k_*$ is a reform supporter, $\ol k_* \in \cC^R$,
and  none of the committee members  $ j \geq \ol k_*$ is transfer recipient and they do not transfer more than their utility.  Moreover,
the post-transfer utilities  $\h \pi(r)  \equiv  u +  r$ are ranked across the coalition of promisers and the coalition of promisees, that is,
\bea
\label{Rank Intensit}
\h \pi_i(r) \le \h \pi_j(r) \q\mbox{for all}\q i\le \ul k_* < \ol k_* \le j.
\eea
\end{lem}


\no{\bf Proof:}  First, since $r_{\ol k_*} < 0$, by Lemma \ref{lem: intermediary step} we have $u_{\ol k_*} +r_{\ol k_*}  \ge 0$, which implies $u_{\ol k_*} > 0$. That is, $\ol k_*\in \cC^R$.
Next, since $r_{\ol k_*} < 0 < r_{\ul k_*}$, by Lemma \ref{lem-switchorder} we have $u_{\ul k_*}  + r_{\ul k_*}  \leq u_{\ol k_*} + r_{\ol k_*}$, and thus  $u_{\ul k_*}   < u_{\ol k_*} $. Then by the ranking condition $u_1\leq u_2 \leq \cds \leq u_I$,  we obtain $\ul k_*<\ol k_*$.

We now prove the statements in \reff{ulolkappa2}. First, the inequality $\ul k_*<\ol k_*$ and the definitions \reff{ulolkappa}  imply that $r_k = 0$ for all $ \ul k^*< k < \ol k_*$. Next, assume by contradiction that  $r_i<0$ for some $i\le \ul k_*$. Since $r_{\ul k_*} > 0$, we must have $i<\ul k_*$. Since the vector $u$ is ranked, we have $\h \pi_i(r) < u_i \le u_{\ul k_*} < \h \pi_{\ul k_*}(r)$. This contradicts Lemma \ref{lem-switchorder}, and thus $r_i\ge 0$ for all $i\le \ul k_*$. Similarly, assume by contradiction that there exists $j\ge \ol k_*$ such that $r_j> 0$. Then we would have $r_{\ol k_*} <0<r_j$ and $ \h \pi_{\ol k_*}(r) < u_{\ol k_*}  \le u_j < \h \pi_j(r)$. This contradicts Lemma \ref{lem-switchorder}, and thus $r_j\le 0$ for all $j\ge \ol k_*$. Moreover, if $r_j < - u_j$ for some $j\ge \ol k_*$, then $\h \pi_j(r) = u_j + r_j < 0$. Since $\ol k_* \in \cC^R$, then $j\in \cC^R$, and thus $r_j < - u_j \le 0$. To sum up, $r_j<0$ and $\h \pi_j(r) <0$. This contradicts Lemma \ref{lem: intermediary step}, so   $r_j \ge - u_j$ for all $j\ge \ol k_*$, and thus \reff{ulolkappa2} holds true.

We finally prove inequalities \reff{Rank Intensit}. Assume by contradiction that $\h \pi_j(r) < \h \pi_i(r)$ for some $i\le \ul k_* < \ol k_* \le j$. By \reff{ulolkappa2} we have $r_j \le 0 \le r_i$. If $r_j < 0 < r_i$, we obtain the contradiction with Lemma \ref{lem-switchorder}. If $r_j=0 < r_i$, we have $r_{\ol k_*} < 0 < r_i$, and since the utility vector is ordered, we have $\h \pi_{\ol k_*} (r) < u_{\ol k_*} \le u_j = \h \pi_j(r) < \h \pi_i(r)$,  contradicting Lemma \ref{lem-switchorder}. Similarly,   if $r_j<0 = r_i$, we have $r_j< 0< r_{\ul k_*}$, and, since the utility vector is ordered, we have $\h \pi_{\ul k_*} (r) > u_{\ul k_*} \ge u_i = \h \pi_i(r) > \h \pi_j(r)$,  also contradicting Lemma \ref{lem-switchorder}. In the last case that $r_i=0=r_j$, we have $r_{\ol k_*} < 0< r_{\ul k_*}$, and hence $\h \pi_{\ol k_*}(r) < u_{\ol k_*} \le u_j = \h \pi_j(r) < \h \pi_i(r) = u_i \le u_{\ul k_*}< \h \pi_{\ul k_*}(r) $. This again contradicts Lemma \ref{lem-switchorder}. In summary, we obtain the desired contradiction  in all the sub-cases, and thus \reff{Rank Intensit} holds true.
\qed

\ms

\no{\bf Proof Proposition \ref{prop-order}:} We now use the lemmas to prove   Proposition \ref{prop-order}. We note that, from the proof below, in all the cases $\ul k_* < k_* \le \ol k_*$.  In this proof, we allow $k_*$ to depend on $r\in  \pmb{\cS \cM}_R \backslash \{ 0\}$. In the supplemental Appendix A, we show that a common $k_*$ can be selected for all $r\in  \pmb{\cS \cM}_R \backslash \{ 0\}$.

We first prove the statements in \reff{order}. Define $ a := \min_{\ol k_* \le j \le I} \h \pi_j(r)$. By  \reff{ulolkappa2} we have $\h \pi_k(r) = u_k$ for $\ul k_* < k < \ol k_*$. If $a \ge  \h \pi_{\ol k_*-1}(r)$, then set $k_* = \ol k_*$. Since the utility vector is ordered,  we have $a \ge \h \pi_{\ol k_*-1}(r) \ge \h \pi_k(r)$ for all $\ul k_* < k < \ol k_*$, and by \reff{Rank Intensit} we have $a\ge \h \pi_i(r)$ for all $i \le \ul k_*$. Moreover, by \reff{ulolkappa2} and  \reff{Rank Intensit} we see that $r_i \ge 0 \ge r_j$ for all $i< k_* \le j$. So $k_* = \ol k_*$ satisfies all the requirements in inequalities \reff{order}. We  next assume $a <  \h \pi_{\ol k_*-1}(r)$. Since  $a \ge \h \pi_{\ul k_*}(r)$, we may set $k_* = \inf\{k > \ul k_*: \h \pi_k (r) > a\}$.  Then $\ul k_* < k_*\le \ol k_*$, and since the utility vector is ordered, we see that $\h \pi_k(r) \le a$ for all $\ul k_* < k < k_*$, and $\h \pi_k(r) \ge \h \pi_{k_*}(r) > a$ for all $k_* \le k < \ol k_*$. Again by \reff{Rank Intensit},  $\h \pi_i(r) \le a \le \h \pi_j(r)$ for all $i< k_* \le j$, and by \reff{ulolkappa2},  $r_i \ge 0 \ge r_j$ for all $i< k_* \le j$. This completes the proof of \reff{order}. Moreover, since $ r\in \cP$, then $\sum_{i<k_*} r_i = \sum_{j\ge k_*} (-r_j)$, and thus we obtain \reff{cTr+-}.
\qed

\ms

\no{\bf Proof of Proposition \ref{prop-dynamic}:} To simplify the exposition, we make two relaxations. First, if a member $i$ is indifferent between $R$ and $S$, we assume they vote for $S$ in the first step of the blocking coalition sequence. Second, we replace strict dominance with a Pareto improvement for a blocking coalition $\cC$, requiring $\hat\pi_i(\tilde{t}) \ge \hat\pi_i(t)$ for all $i \in \cC$, with strict inequality for some $i \in \cC$.  We note that enforcing the original strict dominance would require defining $\epsilon$-equilibria, making the proof more technical.

\no(i) By setting $t(S) \equiv 0$ and $r^t = r$, we have $u + r^t = u + r$. Since $r\in \pmb{\cS \cM}_R$, we have $|\cC|\ge \k$, where $\cC := \{i: u_i + r_i \ge 0\}$. Then $O(t) = R$, and $\hat\pi_i(t) = u_i + r_i \ge 0 = s^{t^0}_i=\hat\pi_i(t^0)$ for all $i\in \cC$. Moreover, since $\sum_{i\in \dbI} (u_i+r_i) = \sum_{i\in \dbI} u_i >0$, there exists $i\in \cC$ such that $u_i + r_i>0$, which implies $\hat\pi_i(t) > \hat\pi_i(t^0)$. Then $\cC$ is a blocking coalition for $t^0$ with corresponding $t$, in our relaxed sense.

\ms
\no (ii) We proceed in two steps. 
{\bf Step 1.}  First, denote $\a:= {\sum_{j\in \ul \cC^R} u_j \over \sum_{j\in \cC^S} |u_j|}>0$ and  define $s\in \cP$ by
$$
s_i = \a u_i<0,~ i\in \cC^S;\q s_i := u_i \ge 0,~ i \in \ul \cC^R;\q  s_i := 0, ~ i\in \cC^R\backslash \ul \cC^R,
$$
There exists $t\in \cA^I$ such that $t(R)\equiv 0$, $s^t = s$ and $t_{ij}(S)>0$ can hold only when $i\in \cC^S$ and $j\in \ul\cC^R$, that is, the promises are made only from members in $\cC^S$ to members in $\ul \cC^R$. We now verify that $\cC^S$ is a relaxed blocking coalition for $t^0$ with the promise profile $t$. First, since $G^S>0$, we have $0\le \sum_{j\in \ul \cC^R} u_j < \sum_{j\in \cC^S} |u_i|$. Then  $\a <1$, and thus $ u_i < s_i$ for $i\in \cC^S$. This implies that
$u^t_i = u_i - s_i <0$ for $ i\in \cC^S$, and $u^t_i = u_i - s_i =0$ for $i\in \ul\cC^R$.
Due to our relaxation, $|\cC^S \cup \ul\cC^R| = \hat\k$ implies $O(t) = S$
Moreover,  $O(t^0)=R$ implies
$\hat\pi_i(t) - \hat \pi_i(t^0) = s_i - u_i  = (\a-1)u_i>0$ for any $i\in \cC^S$. Thus the coalition $\cC^S$  blocks $t^0$.

\no{\bf Step 2.} In this step we show that $\tilde \cC := \cC^R$ is a blocking coalition for $t$. We first let  the members in $\ul\cC^R$ to return $s^t$ to the members in $\cC^S$: $t'(S) := - t(S)$.  Next, noting that $r^t = 0$, we set $ t'(R)$ be such that $r^{t'} =  r$. Denote $\tilde t := t + t'$. Then $r^{\tilde t} - s^{\tilde t} = r$. Since $r\in \pmb{\cS\cM}_R$, then $O(\tilde t) = R$. Moreover, recalling that $O(t) = S$, we have for all $i \in \cC^R$,
\beaa
\hat\pi_i(\tilde t) - \hat\pi_i(t) = u_i + r^{\tilde t}_i - s^t_i = u_i + r_i - s_i = \left\{\ba{lll} u_i+r_i - u_i = r_i \ge 0,~ i\in \ul\cC^R;\\
u_i + r_i \ge 0,~ i\in \cC^R\backslash \ul\cC^R.
  \ea\right.
\eeaa
Since $\sum_{i\in \dbI} (u_i + r_i) = \sum_{i\in \dbI} u_i >0$, by the order preserving property \reff{order} we must have $u_i + r_i > 0$ for some $i\in \cC^R\backslash \ul\cC^R$. This shows that $\tilde t$ Pareto dominates $t$ on $\tilde \cC$. Then, in our relaxed sense, $\tilde \cC$ is a blocking coalition for $t$ with corresponding $\tilde t$.
\qed

Support from the NSF and SSHRC are acknowledged.

\vspace{-0.5cm}
\theendnotes


\newpage

\bibliographystyle{plainnat}
\bibliography{LZ-2023-Aug-06}

\clearpage
\newpage

\begin{center}
\LARGE  Supplemental Web Appendix  for: \\
\textbf{``Unlocking Democratic Efficiency: How Coordinated Outcome-Contingent Promises Shape Decisions''}
\end{center}

\bigskip
\pagenumbering{arabic}
\renewcommand*{\thepage}{SA-\arabic{page}}
\setcounter{section}{0}
\renewcommand{\thesection}{SA.\arabic{section}}
\renewcommand\theequation{SA.\arabic{equation}} \setcounter{equation}{0}
\renewcommand{\theprop}{SA.\arabic{prop}} \setcounter{prop}{0}
\renewcommand{\thelem}{SA.\arabic{lem}} \setcounter{lem}{0}

\thispagestyle{empty} \setcounter{page}{0} \setcounter{section}{0} \newpage

\newpage
\begin{center}
\subsection*{Supplemental Web Appendix A (SA): SM equilibria implications}
\end{center}
In this Appendix, we characterize SM equilibria for all subcases discussed in Section~\ref{Section: Equilibrium promises} of the paper using numerical examples.
In all subsequent discussions, we will work directly with the net transfers $r \in \cP$
since the characterization of strong Nash equilibria from Proposition~\ref{prop: Strong equilibria} only depends on the net transfers $r^t-s^t$. We will use the abuse of notation $O(r)=O(t)$, $\h \pi (r)=\h \pi (t)$ with the understanding that $r^t = r$ and $t(S)\equiv 0$.

\section{The case of ``majority coercion":  $\mid \cC^R \mid < \k$}\label{section: committee with an intense minority}

Assumption \reff{equ: R is efficient} implies that at least one member supports the reform. The case $\k=1$ can be excluded from the proposition, as the reform always passes in this scenario, resulting in no voting inefficiencies. In the proposition, we thus make the assumption that $\kappa \geq 2$.

\begin{prop} \label{prop: Equilibria with an intense minority} [\textbf{Equilibrium transfer promises with majority coercion}]
Consider a committee with weak support for the reform, $\mid\cC^R\mid < \k$ with $\k \geq 2$ resulting in the reform's defeat,  $O(t^0)=S$.  A net transfer  profile $ r \in \cP$ is an SM equilibrium, $r \in \pmb{\cS \cM}_R $, if and only if
\begin{enumerate}

\item The promises are across the aisle: the promise recipients are reform opponents and the promisers are reform supporters. Moreover the total transfer is given by $\cT_{r} = U^S = \sum_{\cC^S}r_i = - \sum_{\cC^R}r_i $.

\item  For all $i\in \cC^R$, the individual rationality constraint  $-u_i \le r_i \le 0$ holds.

\item  All promise recipients are reform opponents, and the net transfers $(r_i)_{i \in \cC^S}$ satisfy:

\begin{enumerate}
\item When $|\cC^R| < \k -1$, each member $i \in \cC^S$ is promised the transfer $r_i = -u_i >0$ just to make her indifferent between the reform and the status quo, $\h \pi_i(r) =0$.

\item When $|\cC^R| = \k -1$, the net transfer of the members of $\cC^S$ are non-negative, $r_i \ge 0$ for all $i \in \cC^S$ and
    the post-transfer utilities of members in $\cC^S$ cannot exceed those of members in $\cC^R$: $\h \pi_i(r) \le \h \pi_j(r) $ for any $i\in \cC^S$ and $j\in \cC^R$.

\end{enumerate}

\end{enumerate}

\end{prop}

In all SM equilibria, the total transfer compensates reform opponents for the aggregate disutility that they experience with the passage of the reform.  The critical voter $k_*$ from Proposition~\ref{prop-order}  is given by $k_*= n+1$, so that the reform supporters are promisers and reform opponents are promisees. The equilibrium leaves multiple ways for the members of the coalition  $\cC^R$  to divide among themselves the total transfer directed towards reform opponents.  Whether there is an additional indeterminacy on the side of the promises recipients depends on the size of the coalition $\cC^R$.

If the coalition $\cC^R$ is short at least $2$ members of being decisive ($|\cC^R| < \k -1$), then the distribution of equilibrium promises among promisees is unique. After receiving the promises, all members of $\cC^S$ are indifferent between voting for or against the reform. 

When the coalition $\cC^R$ is just one member short of being decisive ($|\cC^R| = \k -1$), the transfers to the members of $\cC^S$ are indeterminate. In that case, multiple distributions of promises across the members of the receiving coalition $\cC^S$ can form an equilibrium provided that they satisfy the requirement $3.(b)$ of Proposition~\ref{prop: Equilibria with an intense minority}. The requirement restricts the promises directed to each member of the coalition $\cC^S$ to produce post-transfer utilities that maintain the ordering across the coalitions of promisers and promisees of the utilities prior to the transfer. Changing the ordering of utilities across the coalition $\cC^R$ and $\cC^S$ would create the incentives to engage in additional rounds of promises and contradict the stability requirement of the equilibrium. When $|\cC^R| = \k -1$, the committee adopts the reform after the promises are made, but the vote for the reform may not be unanimous.

\section{Committee with strong support for the reform: $\mid \cC^R \mid \geq \k$}

In this subsection, we study the equilibrium when the committee would adopt the reform in the absence of promises. First,
the following results identify the unique equilibrium promises profile when there are no gains from trade.

\begin{prop} \label{prop: Equilibria in the absence of gain from trade} [\textbf{No promises equilibrium in the absence of gains from trade}]
Consider a committee with more reform supporters than the $\k$-majority requirement, $\mid \cC^R \mid \geq \k$. Assume the gain from trade defined in equation (\ref{eq: gain from trade}) is non-positive, $G^S \leq 0$. Then $ r \equiv 0 $ is the unique SM equilibrium, that is,  $ \pmb{ \cS \cM}_R= \{  0 \} $.
\end{prop}

We now consider the case where the gain from trade $G^S$ is positive with $\mid \cC^R \mid \ge \k$.  Following Section~\ref{section: committee with an intense minority}, we also assume without loss of generality that $\k \ge 2$.

In the following proposition, we show that the intuition in Example \ref{ex: first order} holds more generally and provide conditions under which the coalition of promisers can afford to preempt new rounds of promises and achieve stability.


Specifically, we demonstrate that when $G^S \leq \D U_{\h\k} $, all equilibria involve a total transfer of promises of $G^S$ from the coalition $\cC^R \backslash \underline \cC^R$  to the coalition $\cC^S \cup \underline \cC ^R$. Importantly, in all equilibria, the post-transfer individual utilities remain larger for all members of $\cC^R \backslash \underline \cC^R$ than that of any member of the coalition $\cC^S \cup \underline \cC ^R$. Before stating the proposition, we recall that $\cC^S \cup \underline \cC ^R = \{ 1,..,\h\k \}$ and  $\cC^R  \backslash \underline \cC^R = \{\h\k+1,..,I \}$.

\begin{prop} \label{prop: Equilibria in presence of gain from trade1} [\textbf{Equilibrium promises with first order preemption.}]
Consider a committee in which the support for the reform is larger than the $\k$-majority requirement, $\mid \cC^R \mid \ge \k$ with $\k\ge 2$.
Assume the gain from trade defined in equation (\ref{eq: gain from trade}) satisfies  $0 < G^S \le \D U_{\h\k} $ where $ \D U_{\h\k}$ is defined in equation~(\ref{Delta Ui}). The net transfer profile $r$ is an SM equilibrium,  $ r \in \pmb{\cS \cM}_R$,  if and only if

\begin{enumerate}
\item Members of the coalition $\cC^R\backslash \underline \cC^R$ are promisors subject to individual rationality constraints while members of the coalition $ \cC^S\cup \underline \cC^R$ are promisees:
    \bea
\label{equ: structure of r}
-u_j \le r_j \le 0 \le r_i, \q  \forall i \leq \h\k < j.
\eea

\item The post-transfer utilities of members of the coalition $ \cC^S\cup \underline \cC^R$ cannot exceed those of members of the coalition  $\cC^R\backslash \underline \cC^R$
 \bea
\label{equ: ordering of intensities}
\h \pi_i(r)  \le \h \pi_j(r) , \q  \forall i \leq \h\k  < j.
\eea
\item  The total  transfer induced by $ r$ is given by
\bea
\label{equilibrium1}
\cT_{ r} = G^S = \sum_{i=1}^{\h \k} r_i = - \sum_{j=\h \k +1}^I r_j.
\eea

\end{enumerate}
\end{prop}

Proposition~\ref{prop: Equilibria in presence of gain from trade1} shows that, under the assumption $|\cC^R|\ge \k$ and $0 < G^S \le \D U_{\h\k } $, all equilibria require members of the coalition $\cC^R\backslash \underline \cC^R$ to promise a total transfer of $G^S$ to the members of the coalition $ \cC^S\cup \underline \cC^R$.  In particular, the critical member $k_*$ from Proposition~\ref{prop-order} is given by $k_*= \h\k+1$. 

We now consider the case where $\mid \cC^R \mid \ge \k$ with $\k \geq 2$ and with   $0 \leq \D U_{\h\k }  < G^S$. The following proposition characterizes the SM equilibria in that case.

\begin{prop} \label{prop: Equilibria in presence of gain from trade2 Alternative} [\textbf{ Equilibria with higher-order preemption }]
Consider a committee in which support for the reform is in excess of $\k$-majority requirement, $\mid \cC^R \mid \ge \k$ with $\k \ge 2$.
Assume the gain from trade defined in equation (\ref{eq: gain from trade}) is not affordable to promise, i.e.,  $\D U_{\h\k } < G^S$ where $ \D U_{\h\k}$ is defined in equation~(\ref{Delta Ui}). A  transfer promises profile $r \in \cP$ is an SM equilibrium, if and only if

\begin{enumerate}
  \item  The critical voter $k_*$ from Proposition~\ref{prop-order}  identifies with the $k_*$ defined in equation~(\ref{k*}).

  \item Members of the coalition $\{ k_*,\cds,I\}$ are promisers subject to individual  rationality constraints
   \bea
   \label{rj<0}
      -u_j \leq r_j = -u_j + u_* <0 \q \mbox{for all} \q j \geq k_* \, ,
    \eea
    and experience equal {\it ex post} intensities
    \bea
   \label{ex post intensity}
     \h \pi_j(r)   = u_* >0\q \mbox{for all} \q j \geq k_*.
    \eea

  \item Members of the coalition $\{ 1,\cds,k_*-1\}$ are promisees, and their {\it ex post} intensities cannot exceed those of promisers
    \bea
    \label{vi<vj}
      r_i \geq 0 \q \mbox{ and } \q \h \pi_i(r) \leq \h \pi_j(r)  \equiv  u_*,\q \mbox{for all} \q i<k_*\leq j.
     \eea
  \end{enumerate}

Moreover, equilibrium  promises profiles are indeterminate, and they all generate the common total promises transfer
    \bea
\label{cT* minimal}
\cT_{ r} =\cT_* \mbox{ where } \cT_* := \sum_{j\ge k_*} \big[u_j - u_* \big] > G^S.
\eea
\end{prop}

Proposition~\ref{prop: Equilibria in presence of gain from trade2 Alternative} shows that in all equilibria, members of the promisers coalition promise a transfer that is affine in their utilities as described in equation~\reff{rj<0}. This implies that members with larger intensities promise a larger transfer and this results in an equalized distribution of the post-transfer utilities among the coalition of promisers. Therefore, in all equilibria, the distribution of transfer promises among promisers is unique. This uniqueness is novel and contrasts with the cases covered in Proposition~\ref{prop: Equilibria with an intense minority} and
Proposition~\ref{prop: Equilibria in presence of gain from trade1} where the distribution of transfer promises among promisers was indeterminate. The intuition of this uniqueness is that if the promisers have unequal post-transfer utilities, some of them will become subject to enticement to cast their vote against the reform and this, in turn, contradicts the strong Nash requirement as additional promises are required to preclude additional enticements. On the other hand, the multiplicity associated with the transfer distribution among promisees remains valid as in the cases covered in  Proposition~\ref{prop: Equilibria in presence of gain from trade1}.

Proposition~\ref{prop: Equilibria in presence of gain from trade2 Alternative} also shows that the total transfer is larger than the gain from trade $G^S$ as we illustrated in Example~\ref{ex: second order}. More specifically,  by equation \reff{k*}, the fact that $\hat \k = I-\k+1$, and \reff{eq: gain from trade}, we have
$$
\sum_{j= \h\k +1}^I \big[u_j - u_* \big]= \sum_{j= \h\k +1}^I u_j - (I-\h\k)u_* = \sum_{j= \h\k +1}^I u_j - (\k-1)u_* = \sum_{j= \h\k +1}^I u_j - \sum_{j= 1}^I u_j =-\sum_{j=1}^{ \h\k} u_j = G^S.
$$
Therefore, the aggregate promises from the coalition $\cC^R\backslash\underline \cC^R$ is given by
\bea
\label{eq: gain from trade2}
\sum_{j \in \cC^R\backslash\underline \cC^R} r_j = \sum_{j= \h\k +1}^I \big[u_j - u_* \big] = G^S.
\eea
This shows that members of the coalition $\cC^R\backslash\underline \cC^R$ will promise an aggregate transfer of $G^S$. An aggregate transfer of $G^S$ will be sufficient to achieve an equilibrium when the gain from trade is affordable. However, this is not the case since $\D U_{\h\k } < G^S$ and as a result, some members of  $\cC^R\backslash\underline \cC^R$ will have lower post-transfer utilities than some members of $\underline \cC^R$ and hence become targets for enticement. To preempt this from happening, additional transfer promises are needed from the members of the coalition $\{ k_*,..,\h\k\} \subseteq \underline \cC^R$ with {\it interim} intensities that are larger than those of some members of $\cC^R\backslash\underline \cC^R$. The aggregation of these promises are given by
$
\sum_{j =k_*}^{\h\k}  \big[u_j - u_* \big]>0,
$
 which results in a total transfer $\cT_*  > G^S$.


\section{Proofs  of the  propositions in the Supplemental Appendix A}
\ms
{\bf Proof of Proposition~\ref{prop: Equilibria with an intense minority}:} We start with the following lemma.
\begin{lem}
\label{lem-r}
For any promises profile $r \in \cP$, the following holds:

(i) For any coalition $\cC \subset \dbI$, we have $\cT_{ r} \ge \big|\sum_{i\in \cC} r_i\big|$.

(ii) Consider a coalition $\cC \subset \dbI$ such that  $|\cC| = \h\k$, $u_i+r_i \le u_j + r_j$ for all $i\in \cC$ and $j\notin \cC$.  Then $r \in \pmb{\cS}_R$ if and only if $\sum_{i\in \cC} (u_i +r_i) \ge 0$.
\end{lem}

\no{\bf Proof.}  We first prove the statement (i). Since $r \in \cP$, we have
\beaa
\cT_{\bm{r}} &=& {1\over 2} \Big[ \sum_{i\in \cC} |r_i| + \sum_{i\notin \cC} |r_i|\Big] \ge {1\over 2} \Big[ \big|\sum_{i\in \cC} r_i\big| + \big|\sum_{i\notin \cC} r_i\big|\Big] \\
&=& {1\over 2} \Big[ \big|\sum_{i\in \cC} r_i\big| + \big|-\sum_{i\in \cC} r_i\big|\Big]= \big|\sum_{i\in \cC} r_i\big|.
\eeaa

We next prove the statement (ii).
 First, if $r\in \pmb \cS_R$, since $|\cC| = \h \k $,  by \reff{eq: Strong Nash Characterisation2}
we have $\sum_{i\in \cC} (u_i +r_i) \ge 0$.

We now  assume  $\sum_{i\in \cC} (u_i+r_i)\ge 0$ for $\cC$ satisfying the conditions in part (ii) of Lemma~\ref{lem-r} and prove that $ r \in \pmb \cS_R$. Condition (ii) of Lemma~\ref{lem-r} implies that
\bea \label{equ: majoration}
\min_{j\notin \cC} (u_j+r_j) \ge \max_{i\in \cC} (u_i+r_i) \ge 0.
\eea
For any  coalition $\tilde \cC $ satisfying $|\tilde \cC| \geq \hat\k$, consider the partition of $\tilde \cC$ defined by  $\tilde \cC = \cC_1\cup \cC_2 \cup \cC_3$. Members of the coalition $\cC_1$ belong both to $\cC$ and $\tilde \cC$, that is,  $\cC_1 := \cC\cap \tilde \cC$. The coalition $\cC_2$ is a subset of $ \tilde \cC\backslash \cC$ such that when merged with the coalition $\cC_1$, it forms a coalition with cardinality $\h\k$, that is, $ |\cC_1|+|\cC_2| = \h\k $. Finally, the coalition $\cC_3$ formed by the residual members of $\tilde \cC$ who do not belong to $\cC_1$ or $\cC_2$, that is, $\cC_3:= \tilde \cC\backslash (\cC_1\cup \cC_2)$.
Note that $u_i + r_i \le u_j + r_j$ for all $i\in \cC\backslash \cC_1$ and $j \in \cC_2$ and that, $|\cC\backslash \cC_1| = |\cC_2|$. Thus, the inequality \reff{equ: majoration} implies
\bea \label{equ: majorationBis}
\sum_{j\in \cC_2} (u_j+r_j) \ge \sum_{i\in \cC\backslash \cC_1} (u_i+r_i).
\eea
Since $\cC_3 \cap \cC = \emptyset$, we have $u_k+r_k \ge \max_{i\in \cC} (u_i+r_i) \ge 0$ for all $k\in \cC_3$, and hence
 $ \sum_{k\in \cC_3} (u_k+r_k) \ge 0$.
Using this last inequality and \reff{equ: majorationBis}
yields
\beaa
\sum_{i\in \tilde \cC} (u_i+r_i) &=& \sum_{i\in \cC_1} (u_i+r_i)  + \sum_{j\in \cC_2} (u_j+r_j) +\sum_{k\in \cC_3} (u_k+r_k) \\
&\ge& \sum_{i\in \cC_1} (u_i+r_i)  + \sum_{i\in \cC\backslash \cC_1} (u_i+r_i) + 0  = \sum_{i\in \cC} (u_i+r_i) \ge 0.
\eeaa
Now  it follows from \reff{eq: Strong Nash Characterisation2}
again that $ r \in \pmb \cS_R$.
\qed

\ms

We now prove  Proposition~\ref{prop: Equilibria with an intense minority}. We first show that $\cT_{ r} \ge U^S$ for any net transfer $r\in  \pmb \cS_R$. Indeed,  since $|\cC^R| < \k$, we have $|\cC^S| \geq \h \k$,   then \reff{eq: Strong Nash Characterisation2} 
implies  that $\sum_{\cC^S} (u_i +r_i) \geq 0$. Therefore,
\bea
\label{US1}
\sum_{i\in \cC^S} |r_i|\ge  \sum_{i\in \cC^S} r_i \ge  \sum_{i\in \cC^S} (-u_i) = U^S.
\eea
Then it follows from Lemma \ref{lem-r} (i) that
\bea
\label{eq: stable PPP bound}
\cT_{ r} \ge |\sum_{i\in \cC^S} r_i |\ge U^S, \mbox{ for all }  r \in \pmb \cS_R.
\eea
 We  prove 
the equivalence in Proposition~\ref{prop: Equilibria with an intense minority} by considering two cases separately.

{\bf Case $1$: $|\cC^R |< \k-1$.} We cover first the ``if" direction ($\Leftarrow$), then the ``only if" direction ($\Rightarrow$).

If a promises profile $ r $ satisfies conditions $1.$, $2.$ and  $3.a$ from Proposition~\ref{prop: Equilibria with an intense minority}, then it can be directly checked that $ r\in \pmb \cP$ and $\cT_{ r} = U^S$. Moreover, by construction we see that $u_i + r_i=0$ for $i\in \cC^S$ and $u_j +r_j\ge 0$ for $j\in \cC^R$. Therefore, $\sum_{i \in \cC} (u_i +r_i) \geq 0$ for all coalitions $|\cC| \geq \h \k $ and hence,  by \reff{eq: Strong Nash Characterisation2}
the promises profile $ r$ is a strong Nash equilibrium, that is  $ r \in \pmb \cS_R$. Since $\cT_{ r} = U^S$, inequality (\ref{eq: stable PPP bound}) shows that the promises profile $ r $ achieves the minimum total promises transfer, and thus $ r \in \pmb{\cS \cM}_R$. In particular, we note that the following promises profile $ r$ satisfies $1.$, $2.$ and  $3.a$ from Proposition~\ref{prop: Equilibria with an intense minority}, and hence is an equilibrium with minimum total promises transfer:
\bea
\label{ intense minority-equilibrium}
r_i := -u_i,~ i \in \cC^S;\qq r_j := - {U^S \over U^R} u_j,~ j\in \cC^R.
\eea

We now prove the only if part. That is, we assume $ r\in \pmb{\cS \cM}_R$ and prove that conditions $1.$, $2.$ and  $3.a$ from  Proposition~\ref{prop: Equilibria with an intense minority} hold. Note that
\bea
\label{US2}
\sum_{i\in \cC^R} |r_i| \ge \sum_{i\in \cC^R} (-r_i) =  \sum_{i\in \cC^S} r_i  \ge U^S,
\eea
and combining  \reff{US1} with \reff{US2} gives
\bea
\label{US4}
\cT_{ r} = \frac{1}{2} \sum_{i\in \cC^R} |r_i|  + \frac{1}{2} \sum_{i\in \cC^S} |r_i|\ge \frac{1}{2} \sum_{i\in \cC^R} (-r_i) + \frac{1}{2} \sum_{i\in \cC^S} (r_i) \ge U^S.
\eea
Since $ r \in \pmb{\cS \cM}_R$ has minimum total promises transfer, we must have $\cT_{ r} = U^S$.
Therefore, inequalities \reff{US4}  become equalities, that is,
\beaa
\dis U^S= \cT_{ r} = \frac{1}{2} \sum_{i\in \cC^R} |r_i|  + \frac{1}{2} \sum_{i\in \cC^S} |r_i| = \frac{1}{2} \sum_{i\in \cC^R} (-r_i) + \frac{1}{2} \sum_{i\in \cC^S} (r_i) = U^S.
\eeaa
This in turn implies that inequalities \reff{US1} and \reff{US2} are also equalities:
\beaa
\sum_{i\in \cC^S} |r_i| =  \sum_{i\in \cC^S} r_i = U^S,\q \sum_{i\in \cC^R} |r_i|  =  \sum_{i\in \cC^R} (-r_i) =  U^S.
\eeaa
Then
\beaa
|r_i| = r_i,~ i\in \cC^S;\q |r_i| = -r_i,~ i\in \cC^R;\q \mbox{and}\q \sum_{i\in \cC^S} r_i  = \sum_{i\in \cC^R} |r_i|  =   U^S,
\eeaa
and thus
\bea
\label{US3}
r_i \ge 0 \ge r_j,\q \forall i\in \cC^S, j\in \cC^R,\q\mbox{and} \q \sum_{i\in \cC^S} r_i = U^S = \sum_{j\in \cC^R} (-r_j).
\eea
Moreover, the equality $\sum_{i\in \cC^S} r_i = U^S$ implies that $\sum_{i\in \cC^S} (u_i+r_i) = 0$.

Since $|\cC^S|>\h\k$, then for any $i_0\in \cC^S$, $|\cC^S\backslash \{i_0\}| \geq \h \k $, thus  by \reff{eq: Strong Nash Characterisation2} 
we have $\sum_{i\in \cC^S\backslash \{i_0\}} (u_i + r_i) \ge 0$. This, together with $\sum_{i\in \cC^S} (u_i +r_i) = 0$, implies that $u_{i_0} + r_{i_0}\le 0$, for all $i_0\in \cC^S$.  Combining this with the equality $\sum_{i\in \cC^S} (u_i +r_i) = 0$, we must have $u_i + r_i=0$ or $r_i = -u_i$ for all $i\in \cC^S$.

Finally, for each $j\in \cC^R$, since $|\cC^S\cup\{j\}| \geq \h \k$, then  by \reff{eq: Strong Nash Characterisation2} 
we have $0\le \sum_{i\in \cC^S\cup\{j\}} (u_i +r_i) = u_j  + r_j$. That is, $r_j \ge - u_j$ for all $j\in \cC^R$. To summarize, we have proven that, in addition to the relations \reff{US3}, $r_i = -u_i$ for all $i \in \cC^S$ and $-u_j \leq r_j \leq 0$ for all $j\in \cC^R$. Thus, conditions $1$ and  $2.a$ from Proposition~\ref{prop: Equilibria with an intense minority} hold. This concludes the proof for the case $| \cC^R| < \k -1$.

{\bf Case $2$: $|\cC^R | = \k-1$.} 
 Notice that in this case, $|\cC^S| = \h \k$. First, if  $ r$ satisfies conditions $1$ and $2.b$ of Proposition \ref{prop: Equilibria with an intense minority}, then it can be checked as in Step 1 that $ r \in \cP$, $\cT_{ r} = U^S$, $\sum_{i\in \cC^S} (u_i +r_i) =0$, and $r_j \ge -u_j$ for all $j\in \cC^R$. Moreover, by condition $2.b$ of  Proposition \ref{prop: Equilibria with an intense minority} and property (ii) of Lemma \ref{lem-r}, we see that $r\in  \pmb \cS_R$. Since $\cT_{r} = U^S$, the net transfer $ r$ minimizes the total promises transfer and hence $ r \in \pmb{\cS \cM}_R$. In particular, we note that the promises profile $ r$ constructed in \reff{ intense minority-equilibrium} satisfies conditions $1$ and $2.b$ of of Proposition \ref{prop: Equilibria with an intense minority} and hence it belongs to $ \pmb{\cS \cM}_R$. In contrast to the case where $|\cC^S| < \k -1$ where all members of the coalition $\cC^S$ become indifferent between $R$ and $S$ after receiving the equilibrium transfer promises, we note that in the case $|\cC^R| = \k -1$ it is possible that $u_i +r_i>0$ for some $i\in \cC^S$.

We now prove the only if part. We assume $r\in \pmb{\cS \cM}_R$ and show that conditions $1.$, $2.$ and $3.b$ of Proposition \ref{prop: Equilibria with an intense minority} hold true. By the same arguments in Step $1$ of this proof, we see that \reff{US3} still holds true. Moreover, for any $i_0 \in \cC^S$ and $j_0\in \cC^R$, note that $|(\cC^S\backslash \{i_0\} ) \cup \{j_0\} | \geq \h \k $, then  by \reff{eq: Strong Nash Characterisation2} 
we have
\beaa
0\le \sum_{i\in (\cC^S\backslash \{i_0\} )\cup \{j_0\}} (u_i+r_i) &=& \sum_{i\in \cC^S} (u_i+r_i) - (u_{i_0}+r_{i_0}) + (u_{j_0}+r_{j_0})\\ &=&  (u_{j_0}+r_{j_0}) - (u_{i_0}+r_{i_0}) .
\eeaa
Thus, $ u_{i_0}+r_{i_0} \leq  u_{j_0}+r_{j_0}$ for all $i_0 \in \cC^S$ and $j_0\in \cC^R$ and hence, Condition $3.(b)$ of Proposition \ref{prop: Equilibria with an intense minority} is satisfied.  With the exception of the inequality $- u_j \le r_j$ for $j \in \cC^R$, all other properties in Condition $1.$ and $2.$ from Proposition \ref{prop: Equilibria with an intense minority} are implied by \reff{US3}.
 To prove this last property, note that if $\sum_{i\in \cC^S} (u_i +r_i)= 0$, then there exists $i_0\in \cC^S$ such that $u_{i_0} + r_{i_0}\ge 0$. Condition $2.$ from Proposition \ref{prop: Equilibria with an intense minority} implies that $u_j +r_j\ge u_{i_0} +r_{i_0} \ge 0$ for any $j\in \cC^R$, and thus $r_j \ge -u_j$, for all $j\in \cC^R$.
\qed


\ms

\no {\bf Proof of Proposition~\ref{prop: Equilibria in the absence of gain from trade}:} Note that in this case $\sum_{i=1}^{\h\k} u_i = - G^S \ge 0$. It can be verified that the conditions in Lemma \ref{lem-r} (ii) for $\cC = \{1, \cds, \h\k\}$ and $ r=  0$ hold. Then  $ 0 \in \pmb{\cS }_R$. This degenerate transfer $r^{t^0}$ has zero total transfers and hence it is unique, $ \pmb{\cS \cM}_R= \{ \pmb 0 \} $.
\qed

\ms
\no {\bf Proof of Proposition~\ref{prop: Equilibria in presence of gain from trade1}:} The proof proceeds in four steps. In the first step, we show that  $\cT_{ r} \ge G^S$ for all $r \in \pmb \cS_R$. In the second step, we prove the "if" part of  Proposition~\ref{prop: Equilibria in presence of gain from trade1}. In the third step, we give an example of across the aisle equilibrium promise that satisfies $\cT_{ r} = G^S$. The example is used in the fourth and last step where we prove the "only if" part of Proposition~\ref{prop: Equilibria in presence of gain from trade1}.

{\bf Step 1.} We first show that $\cT_{ r} \ge G^S$ for any $ r\in \pmb{\cS }_R$.
Since $|\cC^S\cup \underline \cC^R | \geq \h \k $,  by  
\reff{eq: gain from trade}  and \reff{eq: Strong Nash Characterisation2} we have
\bea
\label{GS1}
0\le  \sum_{i\in   \cC^S\cup \underline \cC^R} [u_i + r_i] = - G^S + \sum_{i\in   \cC^S\cup \underline \cC^R} r_i.
\eea
Then by Lemma \ref{lem-r} (i) we have $\cT_{ r} \ge | \sum_{i\in   \cC^S\cup \underline \cC^R} r_i | \geq \sum_{i\in   \cC^S\cup \underline \cC^R} r_i \ge G^S$.

{\bf Step 2.} We next prove the if part: assume conditions \reff{equ: structure of r}, \reff{equ: ordering of intensities} and \reff{equilibrium1} hold and prove that $ r \in \pmb{\cS \cM}_R$. Let $ r\in \cP$ satisfy  \reff{equ: structure of r}, note that $r_k \ge -u_k$ or $u_k +r_k\ge 0$ for all $k\in \cC^R\backslash \underline \cC^R$, and by the calculation in \reff{GS1} we see that $\sum_{i\in   \cC^S\cup \underline \cC^R} \h \pi_i(r)  \ge 0$. These observations, together with condition \reff{equ: ordering of intensities} show that all conditions required in Lemma \ref{lem-r} (ii) are fulfilled, and therefore we have $ r \in  \pmb{\cS}_R$. Condition \reff{equilibrium1} and Step $1$ of this proof show that in addition to being a strong Nash equilibrium, the net transfer $r$ achieves the minimal total promises transfer $\cT_{ r} = G^S$, and thus $ r \in \pmb{\cS \cM}_R$.

{\bf Step 3.} In this step we construct an equilibrium $ r\in \pmb{\cS \cM}_R$. Consider the following across the aisle net transfer  $ r$:
\bea
\label{equilibrium1-example}
\!\!\! r_i := -{G^S\over U^S} u_i,~ i \in \cC^S;~~ r_j := 0,~ j \in \underline \cC^R;~~ r_k := -{G^S\over {\D U_{\h\k}}}[u_k - u_{\h\k}],~ k\in \cC^R\backslash \underline \cC^R.
\eea
We can show directly that the net transfer defined in \reff{equilibrium1-example} satisfies conditions \reff{equ: structure of r}, \reff{equ: ordering of intensities} and, \reff{equilibrium1}.To see this, first observe that $ r \in \cP$ since, using \reff{equ: aggregate intensity of preferences for R or for S} and \reff {Delta Ui}, it can be checked that  $\sum_\mathbb{I} r_i =0$. Noticing that $u_{\h\k} \ge 0$, $G^S \le U^S$, and $G^S \le \D U_{\h\k }$, we have for any $i\in \cC^S$, $j\in \underline \cC^R$, and $k\in \cC^R\backslash \underline \cC^R$,
\beaa
r_i \ge 0,\q r_j =0,\q  r_k \le 0;\qq r_k \ge  -{G_s\over U_{\h\k}}u_k\ge -u_k \, ,
\eeaa
and therefore $ r$ satisfies condition \reff{equ: structure of r}. Moreover, we have,
\beaa
& \h\pi_i(r) = [1- {G^S\over U_s}] u_i < 0 \le u_{\h\k},\q \h \pi_j(r) = u_j \le u_{\h\k},\\
& \h \pi_k(r) = u_k -{G_s\over \D U_{\h\k}}[u_k - u_{\h\k}] \ge u_k -[u_k - u_{\h\k}] =u_{\h\k};
\eeaa
and thus condition \reff{equ: ordering of intensities} is satisfied. Finally,
\beaa
& \sum_{i\in \cC^S\cup \underline \cC^R} r_i = {G^S\over U^S} \sum_{i\in \cC^S} [-u_i] = G^S,\\
& \sum_{k\in \cC^R\backslash \underline \cC^R} r_k =- {G_s\over \D U_{\h\k}}\sum_{k\in \cC^R\backslash \underline \cC^R} [u_k - u_{\h\k}] = - {G_s\over \D U_{\h\k}} \D U_{\h\k} = - G^S;
\eeaa
and hence condition \reff{equilibrium1} is satisfied. To sum up, the net transfer $ r$ satisfy conditions  \reff{equ: structure of r}, \reff{equ: ordering of intensities} and, \reff{equilibrium1}, and therefore it is an SM equilibrium. 

To sum up, using the result in Step $2$ of this proof, shows that the promises profile $ r$ is an SM equilibrium.

{\bf Step 4.} We now prove the only if part. Let $ r\in \pmb{\cS \cM}_R$ and prove that $ r$ satisfies conditions  \reff{equ: structure of r}, \reff{equ: ordering of intensities} and, \reff{equilibrium1}. Applying Step $1$ of this proof shows that, since $ r \in \pmb{\cS }_R$, we have $\cT_{ r} \ge G^S$. Recall \reff{GS1} and note that
\bea
\label{equ: Inequality for the left coalition}
&\dis \sum_{i\in   \cC^S\cup \underline \cC^R} |r_i| \ge \sum_{i\in   \cC^S\cup \underline \cC^R} r_i \ge G^S;\\
\label{equ: Inequality for the right coalition}
&\dis\sum_{k\in  \cC^R\backslash \underline \cC^R} |r_k| \ge \sum_{k\in  \cC^R\backslash \underline \cC^R} (-r_k) = \sum_{i\in  \cC^S\cup \underline \cC^R} r_i \ge G^S.
\eea
Combining \reff{equ: Inequality for the left coalition} and \reff{equ: Inequality for the right coalition} gives
\bea
\label{equ: Inequality for the total payment}
\cT_{ r} =\frac{1}{2} \sum_{i\in   \cC^S\cup \underline \cC^R} |r_i| + \frac{1}{2}\sum_{k\in  \cC^R\backslash \underline \cC^R} |r_k| \ge
\frac{1}{2} \sum_{i\in   \cC^S\cup \underline \cC^R} r_i + \frac{1}{2}\sum_{k\in  \cC^R\backslash \underline \cC^R}( -r_k) \ge G^S.
\eea
Since $\cT_{ r} \ge G^S$ for any equilibrium net transfer, and since the equilibrium net transfer defined in \reff{equilibrium1-example} achieves the total promises transfer $G^S$, it must be that  $\cT_{ r} = G^S$ for any $r \in \pmb{\cS \cM}_R$. When $\cT_{ r} = G^S$, inequalities \reff{equ: Inequality for the total payment} become equalities:
\beaa
G^S=\cT_{ r} =\frac{1}{2} \sum_{i\in   \cC^S\cup \underline \cC^R} |r_i| + \frac{1}{2}\sum_{k\in  \cC^R\backslash \underline \cC^R} |r_k| =
\frac{1}{2} \sum_{i\in   \cC^S\cup \underline \cC^R} r_i + \frac{1}{2}\sum_{k\in  \cC^R\backslash \underline \cC^R} (-r_k) = G^S.
\eeaa
This, in turn, implies inequalities \reff{equ: Inequality for the left coalition} and \reff{equ: Inequality for the right coalition} are also equalities:
\beaa
\sum_{i\in   \cC^S\cup \underline \cC^R} |r_i| = \sum_{i\in   \cC^S\cup \underline \cC^R} r_i = G^S,\q \sum_{j\in  \cC^R\backslash \underline \cC^R} |r_j| = \sum_{j\in  \cC^R\backslash \underline \cC^R} (-r_j) =  G^S.
\eeaa
Then similarly to the approach used to prove \reff{US3}, we deduce that $|r_i| = r_i$ for all $i \in \cC^S\cup \underline \cC^R$ and $|r_j| = -r_j$ for all $j \in \cC^R\backslash \underline \cC^R$. Hence
\beaa
r_j \le 0 \le r_i,\q \forall i\in \cC^S\cup \underline \cC^R, j\in \cC^R\backslash \underline \cC^R.
\eeaa
Recalling that $\cC^S \cup \underline \cC ^R = \{ 1,..,\h\k \}$ and  $\cC^R  \backslash \underline \cC^R = \{\h\k+1,..,I \}$, we see that the last equation is equivalent to
\beaa
r_j \le 0 \le r_i,\q \forall i \leq \hat \k < j.
\eeaa
Moreover,
\beaa
\cT_{ r}= \sum_{i\in \cC^S\cup \underline \cC^R} r_i \equiv  \sum_{i=1}^{\h \k} r_i = \sum_{j\in \cC^R\backslash \underline \cC^R} (-r_j) \equiv  \sum_{j=\h \k +1}^{I} (-r_j) = G^S.
\eeaa
Using \reff{eq: gain from trade}, observe further that
\beaa
\sum_{i\in   \cC^S\cup \underline \cC^R} (u_i + r_i)  = \sum_{i\in   \cC^S} u_i +  \sum_{i\in  \underline \cC^R} u_i + \sum_{i\in   \cC^S\cup \underline \cC^R}  r_i =  - U^S + \underline U^R + G^S =0.
\eeaa
To verify the individual rationality constraint $-u_j \leq r_j$ for any $j \in \cC^R\backslash \underline \cC^R$, observe that  $|( \cC^S\cup \underline \cC^R) \cup \{j\} | \geq \h \k n$, then  by \reff{eq: Strong Nash Characterisation2}
we have
\beaa
0 \le \sum_{i\in ( \cC^S\cup \underline \cC^R) \cup \{j\}} (u_i+r_i) =\sum_{i\in  \cC^S\cup \underline \cC^R} (u_i+r_i) + (u_j +r_j) =  u_j + r_j,
\eeaa
which implies $r_j \ge - u_j$. So far, we have shown that the net transfer $ r$ satisfies  \reff{equ: structure of r} and \reff{equilibrium1}.

All that's left to prove is that \reff{equ: ordering of intensities} also holds. For any $j\in  \cC^S\cup \underline \cC^R$ and $k\in  \cC^R\backslash \underline \cC^R$, note that $|(\cC^S\cup \underline \cC^R \cup \{k\}) \backslash \{j\}| = \h\k$, then by \reff{eq: Strong Nash Characterisation2}
we have
\beaa
~~ 0\le \!\!\! \sum_{i\in  (\cC^S\cup \underline \cC^R \cup \{k\}) \backslash \{j\}} \!\!\! (u_i +r_i) = \!\!\! \sum_{i\in  \cC^S\cup \underline \cC^R} \!\!\! (u_i +r_i) + (u_k +r_k) - (u_j +r_j) = \h \pi_k(r) - \h \pi_j(r).
\eeaa
Thus, $\h \pi_j(r) \le \h \pi_k(r)$ for all $j\in  \cC^S\cup \underline \cC^R$ and $k\in  \cC^R\backslash \underline \cC^R$, and the net transfer $r$ satisfies \reff{equ: ordering of intensities}, which completes Step 4 and thus concludes the proof.
\qed


\ms
\no {\bf Proof of Proposition~\ref{prop: Equilibria in presence of gain from trade2 Alternative}:} We start the proof with two preliminary lemmas.

\begin{lem}
\label{lem-kstar}
Consider a committee operating under the $\k$-majority rule with $| \k | \geq 2$ and such that $G^S >\D U_{\h \k }$. The following statements hold:

(i)  Recalling the utility $u_*$ defined in \reff{k*}, we have
\bea
\label{u*}
  u_* < u_{\h \k}.
\eea

(ii) There exists a unique group member $k_* \in \underline \cC^R$, i.e., $n < \k_* \leq \h \k$, such that inequalities \reff{k*} hold, that is,
$
u_{k_* -1}  \le u_* < u_{k_*}.
$

(iii) The constant $\cT_*$ defined in \reff{cT* minimal} satisfies the inequality
\bea
\label{cT*>GS}
\cT_* > G^S.
\eea
\end{lem}

\no{\bf Proof.} To prove that (i) holds, observe that
\bea
\label{GS vs D U}
\dis G^S - \D U_{\h \k} &=&  \sum_{i = \h \k +1}^{I} [u_i - u_* ] -  \sum_{i = \h \k +1}^{I} [u_i - u_{\h\k} ] \nonumber\\
\dis &=& (I-\hat\k) [ u_{\h \k} -u_*] = (\k -1) [ u_{\h \k} -u_*].
\eea
where we have used \reff{eq: gain from trade2}, the definition of $\D U_{\h \k}$ in \reff{Delta Ui}, and the relation $\h \k = I - \k +1$.
Since $G^S > \D U_{\hat \k}$, this  establishes inequality \reff{u*} in part (i) of Lemma~\ref{lem-kstar}.

By the relations  \reff{k*}, and \reff{u*},  we have $u_n < 0 < u_* <  u_{\h\k}$. Recalling that the utilities $u_i$ are ordered, there exists a unique $k_* \in  \underline \cC^R $, i.e., $n < k_* \leq \h\k$, such that the inequalities \reff{k*} hold.  Indeed, $k_* := \min k$ such that $u_k > u_*$.
This proves part (ii) of Lemma~\ref{lem-kstar}.

Finally, observe that the total promises transfer $\cT_*$ defined in  \reff{cT* minimal} satisfies
\bea
\cT_* \equiv \sum_{j = k_*}^{\h \k} [u_j - u_*] + \sum_{j = \h \k +1}^I [u_j - u_*] >   \sum_{j = \h \k +1}^I [u_j - u_*] = G^S,
\label{last equation}
\eea
where the first inequality is due to  \reff{k*}, and the last equality is due to \reff{eq: gain from trade2}.
This establishes inequality \reff{cT*>GS} in part (iii) of Lemma~\ref{lem-kstar}. \qed

\begin{lem}
\label{lem-example of promises}
Consider the across the aisle net transfer profile $ r$ defined by
\bea \label{Example of equilibrium promises}
\!\!\! r_i := -{\cT_*\over U^S} u_i,~ \forall i \leq n;\q r_k := 0,~ \forall n< k < k_* ;\q r_j := -[u_j - u_*],~ \forall j\geq k_*,
\eea
where we recall that $k_*$ is defined in part (ii) of Lemma~\ref{lem-kstar}.
Then the net transfer $r$ is zero sum, $ r \in \cP$ and satisfies the conditions \reff{rj<0}-\reff{cT* minimal} of Proposition~\ref{prop: Equilibria in presence of gain from trade2 Alternative}. Moreover, the net transfer profile $ r$ defined in \reff{Example of equilibrium promises} is a strong equilibrium.
\end{lem}
\no{\bf Proof.} First, $ r \in \cP$ since
\beaa
\sum_{i\in \dbI} r_i
= {\cT_*\over U^S}\sum_{i = 1}^n (-u_i) - \sum_{j = k_*}^I [u_j - u_*] = \cT_* - \cT_* =0.
\eeaa
Second, the statements in \reff{rj<0}, \reff{ex post intensity} and \reff{cT* minimal} can be directly checked from the definition of the net transfer $ r $ given in \reff{Example of equilibrium promises}.

Next, to prove \reff{vi<vj}, we need first to prove the preliminary result that $0< \cT_* < U^S$. Using \reff{last equation}, observe that
\beaa
U^S - \cT_* = U^S  - \sum_{j =k_*}^{\h \k} [u_j - u_{*}] - G^S.
\eeaa
Using the definition of $G^S$ given in \reff{eq: gain from trade} yields
\bea
\label{US vs cT}
U^S - \cT_* = \underline U^R  - \sum_{j =k_*}^{\h \k} [u_j - u_{*}] = \underline U^R  - \sum_{j =k_*}^{\h \k} u_j + (\h\k - k_*+1)u_*.
\eea
Since $k_* \in \underline \cC^R$, we have $n<k_* \leq \h \k$ and hence $\underline U^R  - \sum_{j =k_*}^{\h \k} u_j \geq 0$. Moreover, $n<k_* \leq \h \k$ also implies that $(\h\k - k_*+1)u_*>0$. Thus, using \reff{US vs cT} gives $0 < \cT_* < U^S$.

Now we are prepared to prove \reff{vi<vj}. For  $i \le n$, $n < k < k_*$, and $j \ge k_*$, we have
\beaa
& \h \pi_i(r) =u_i + r_i = [1-{\cT_*\over U^S}] u_i \le 0;\\
& \h \pi_k(r)=u_k + r_k = u_k \ge 0;\q \h \pi_j(r)=u_j + r_j = u_*>0,
\eeaa
where the first inequality is implied by $0 < \cT_* < U^S$.  Thus, we see that the condition \reff{vi<vj} is satisfied for the net transfer profile $ r$ defined in \reff{Example of equilibrium promises}.

Finally, to show that $ r$ is a strong equilibrium, set $\cC := \cC^S\cup \underline \cC^R$ in Lemma \ref{lem-r} (ii) so that $|\cC| = \h\k$. Using condition \reff{vi<vj} and observing that $u_j + r_j = u_*$ for all $j \ge k_*$ shows that we have $u_i + r_i \le u_j+ r_j$ for all $i\in \cC$ and $j\notin \cC$.

Moreover, since $ r \in \cP$, by \reff{u*} we have
\beaa
\sum_{i\in \cC}(u_i+r_i) &=& \sum_{i\in \cC^S\cup \underline \cC^R} u_i - \sum_{j\in  \cC^R\backslash\underline \cC^R} r_j =  \sum_{i\in \cC^S\cup \underline \cC^R} u_i - \sum_{j\in  \cC^R\backslash\underline \cC^R} [-u_j +u_* ]\\
&=& \sum_{i\in \dbI} u_i - (\k-1) u_* =0.
\eeaa
Then by applying Lemma \ref{lem-r} (ii), we see that $ r\in \pmb \cS_R$.
\qed



\ms
\no{\bf Proof of Proposition~\ref{prop: Equilibria in presence of gain from trade2 Alternative}:} We proceed in two steps.

{\bf Step 1.} In this step, we prove the only if part: we fix an equilibrium net transfer profile $ r\in \pmb \cS_R$, and show that $ r$ satisfies conditions \reff{rj<0}-\reff{cT* minimal} of Proposition~\ref{prop: Equilibria in presence of gain from trade2 Alternative}. Notice that Proposition~\ref{prop-cEexistence} shows that the set $\pmb \cS_R$ is not empty, and thus it is possible to select a net transfer profile from $\pmb \cS_R$.

Recall Lemma \ref{lem-overline k underline k} and let $\tilde k_*\in \cC^R$ satisfy
 \bea \label{bounds on k_*}
 \underline k_* \equiv \max \{ i: r_i >0\} < \tilde k_* \leq \min \{ i: r_i <0\} \equiv \overline k_*,
 \eea
 and the requirements \reff{order} (In this proof, we reserve the notation $k_*$ for the requirement \reff{k*} instead of the requirements \reff{order}), that is, $-u_j \le r_j \le 0 \le r_i$  and $\h \pi_i(r) \le \h \pi_j(r)$ for all $i < \tilde k_* \le j$. Moreover, we assume $\tilde k_*$ is the largest one satisfying the requirements, and thus
\bea
\label{rtildek*}
\mbox{either}~ r_{\tilde k_*} < 0,\q  \mbox{or} \q \h \pi_{\tilde k_*}(r) >  \min_{j> \tilde k_*} ,\h \pi_j(r)
\eea
because otherwise $\tilde k_*+1$ would also satisfy the desired requirements. To see this, assume the opposite of statement \reff{rtildek*} holds, that is, $r_{\tilde k_*} \ge 0$ and $\h \pi_{\tilde k_*}(r) \le  \min_{j> \tilde k_*} \h \pi_j(r)$. Condition \reff{order} implies that $r_{\tilde k_*} \le 0$, and hence the assumption $r_{\tilde k_*} \ge 0$ implies that $r_{\tilde k_*} = 0$. Thus, we have $\underline k_* < \tilde k_* +1 \le \overline k_*$ and, $-u_j \le r_j \le 0 \le r_i$ for all $i < \tilde k_* +1\le j$. Finally, the assumption $\h \pi_{\tilde k_*}(r) \le  \min_{j> \tilde k_*} \h \pi_j(r)$ implies $\h \pi_i(r) \le \h \pi_j(r)$ for all $i < \tilde k_* +1 \le j$. Hence $\tilde k_*+1$ also satisfies the requirements \reff{bounds on k_*} and \reff{order}.

{\bf Step 1.1.} We first show that $\tilde k_* \le k_*$. Since $k_* \le \h\k$, this implies $\tilde k_* \le \h\k$. Assume by contradiction that $\tilde k_* > k_*$.
Then by  \reff{order} we have
\beaa
u_j + r_j \ge u_{k_*} + r_{k_*} \ge u_{k_*},\q \mbox{for all}\q j \ge \tilde k_*.
\eeaa
Moreover, in the case  $\tilde k_* > \h\k$, recalling that the utilities $u_i$ are ordered, we also have,
\beaa
 u_j + r_j \ge u_j \ge u_{k_*},\q \mbox{for all}\q \h\k\le j < \tilde k_*.
\eeaa
So in all the cases we have $u_j + r_j \ge u_{k_*}$ for all $j\ge  \h\k$.
This, in turn, implies,
\beaa
\sum_{j\in \cC^R\backslash \underline \cC^R} r_j  \ge  \sum_{j > \h\k} [  u_{k_*} - u_j ] = - \D U_{k_*} > - G^S.
\eeaa
Thus, since $ r \in \cP$,
\beaa
\sum_{i\in \cC^S \cup \underline \cC^R} [u_i + r_i] = \sum_{i\in \cC^S \cup \underline \cC^R} u_i  - \sum_{j\in \cC^R\backslash \underline \cC^R} r_j < -G^S + G^S =0.
\eeaa
This contradicts  \reff{eq: Strong Nash Characterisation2}, which is required here because $r\in \pmb \cS_R$ and $ |\cC^S \cup \underline \cC^R|= \h \k$.

{\bf Step 1.2.}  In this step, we prove that the post-transfer utilities of the members $j = \tilde k_*,\cds,I$ are equal:
\bea
\label{equilibrium2-claim2}
\h \pi_{\tilde k_*}(r) = \cds = \h \pi_{I}(r).
\eea
Consider for notational convenience, the order statistics of $\{\h \pi_j(r)\}_{\tilde k_* \le j \le I}$:
\beaa
\h \pi_{l_{\tilde k_*}} (r) \le \cds\le \h \pi_{l_I}(r),
\eeaa
 where $\{l_{\tilde k_*},\cds, l_{I}\}$ is a permutation of $\{\tilde k_*, \cds, I\}$.

To simplify the exposition, we proceed in two substeps. First, we show that $\h \pi_{l_{\tilde k_*}}(r) = \cds = \h \pi_{l_{\h\k+1}}(r)$. Second, we show that the equality $\h \pi_{l_{\tilde k_*}} (r)= \cds = \h \pi_{I}(r)$ is also true.

 {\bf Step 1.2.1.} In this sub-step we show that
\bea
\label{equilibrium2-claim1}
\h \pi_{l_{\tilde k_*}} (r)=  \h \pi_{l_{\h\k+1}}(r) \q\mbox{and hence}\q \h \pi_{l_{\tilde k_*}}(r) = \cds = \h \pi_{l_{\h\k+1}}(r).
\eea
Notice that we are considering the term $l_{\h\k+1}$, rather than $l_{\h\k }$.

Assume by contradiction that
\bea
\label{equilibrium2-claim1-contradiction}
\h \pi_{l_{\tilde k_*}} (r)< \h \pi_{l_{\h\k+1}}(r).
\eea
We claim that
\bea
\label{rk*<0}
r_{l_{\tilde k_*}} <0.
\eea
Indeed, recall \reff{rtildek*}. In the case  $r_{\tilde k_*}<0$, by the definition of order statistics and recalling that the utilities $u_i$ are ordered, we have
\beaa
u_{l_{\tilde k_*}} + r_{l_{\tilde k_*}} \le u_{\tilde k_*} + r_{\tilde k_*} < u_{\tilde k_*}  \le u_{l_{\tilde k_*}},
\eeaa
which implies \reff{rk*<0}. In the  case  $r_{\tilde k_*}=0$ and $\h \pi_{\tilde k_*} (r) >  \min_{j> \tilde k_*} \h \pi_j(r)$, by the ordering of the utilities $u_i$ again, we have
\beaa
u_{\tilde k_*}=\h \pi_{\tilde k_*}(r) >  \min_{j\ge \tilde k_*} \h \pi_j(r) =\h \pi_ {l_{\tilde k_*}} (r)=  u_ {l_{\tilde k_*}} + r_ {l_{\tilde k_*}} \ge u_{\tilde k_*} + r_ {l_{\tilde k_*}},
\eeaa
implying  \reff{rk*<0} again.

Clearly $r_i >0$ for some $i < \tilde k_*$, and assume without loss of generality that $r_1>0$. We now modify $ r$ as follows: for some $\e>0$ small,
\beaa
\tilde r_1 = r_1 - \e > 0,\q \tilde r_{l_{\tilde k_*}} = r_{l_{\tilde k_*}} + \e < 0,\q \mbox{and}\q \tilde r_i = r_i\q\mbox{for all}~ i\neq 1, l_{\tilde k_*}.
\eeaa
Set $\tilde \cC \!=\! \{1,\cds, \tilde k_*-1\}\cup\{l_{\tilde k_*}, \cds, l_{\h\k}\}$ and notice that $|\tilde\cC|=\h\k$. Note that, for $\e>0$ small enough,
\beaa
& u_1 + \tilde r_1  < u_1 + r_1 \le  \h \pi_{l_{\h\k+1}}(r) ;\qq  u_{l_{\tilde k_*}} + \tilde r_{l_{\tilde k_*}}  = \h \pi_{l_{\tilde k_*}}(r) + \e < \h \pi_{l_{\h\k+1}}(r);\\
& u_i + \tilde r_i = u_i + r_i \le \h \pi_{l_{\h\k+1}}(r),\q i \in \tilde \cC\backslash \{1, l_{\tilde k_*}\};\q  u_j + \tilde r_j = u_j + r_j \ge \h \pi_{l_{\h\k+1}}(r),\q j\notin \tilde\cC.
\eeaa
Thus $u_i + \tilde r_i \le \h \pi_{l_{\h\k+1}}(r) \le u_j + \tilde r_j$ for all $i\in \tilde\cC$ and $j\notin \tilde\cC$. Note further that, since $ r\in \pmb{\cS \cM}_R \subset \pmb \cS_R$,
\beaa
 \sum_{i\in \tilde \cC} [u_i + \tilde r_i] = \sum_{i\in \tilde \cC} [u_i +  r_i] + \Big[(\tilde r_1 - r_1) + (\tilde r_{l_{\tilde k_*}} - r_{l_{\tilde k_*}})\Big] = \sum_{i\in \tilde \cC} [u_i +  r_i]  \ge 0.
\eeaa
Then by Lemma \ref{lem-r} (ii) we have $\tilde r \in \pmb \cS_R$. However, as in the last part of the proof for Lemma \ref{lem-switchorder}, we have $
\cT_{\tilde { r}} =   \cT_{ r} -\e < \cT_{ r} $.
This contradicts the assumption that $ r\in \pmb{\cS \cM}_R$ has the minimum total promises transfer. Therefore, \reff{equilibrium2-claim1} holds true.

{\bf Step 1.2.2.} Now we proceed to prove \reff{equilibrium2-claim2}.
Assume by contradiction that, for the order statistics in the previous step,
\beaa
\h \pi_{l_{\tilde k_*}}(r) = \cds = \h \pi_{l_{k_2}}(r) < \h \pi_{l_{k_2+1}}(r) \le \cds\le \h \pi_{l_I}(r),\q\mbox{for some}~ \h\k +1 \le k_2 < I.
\eeaa
 First, by \reff{rk*<0} we also have
\bea
\label{rk*<0all}
r_{l_j} <0,\q\mbox{for all}\q j = \tilde k_*,\cds, k_2.
\eea
Again assume $r_1>0$.
We then modify $r$ as follows: for $\e>0$ small,
\beaa
&\dis \hat r_1 = r_1 - [\h\k - \tilde k_* + 1]\e > 0,\q \hat r_{l_j} = r_{l_j} + \e < 0,~ j = \tilde k_*,\cds, k_2;\\
&\dis \hat r_{l_{k_2+1}} = r_{l_{k_2+1}}  - [k_2 - \h\k]\e<0;\q \hat r_i = r_i\q\mbox{for all other $i$}.
\eeaa
One can check that $\pmb r \in \cP$:
\beaa
\sum_{i\in \dbI}\hat r_i = \sum_{i\in \dbI} r_i - [\h\k - \tilde k_* + 1]\e + \sum_{j=\tilde k_*}^{k_2} \e - [k_2 - \h\k]\e=0.
\eeaa

Similarly to Step $1.2.1$, we see that, for all $i < \tilde k_*$ and  $j > k_2+1$,
\beaa
u_i + \hat r_i \le u_{l_{\tilde k_*}} + \hat r_{l_{\tilde k_*}} = \cds = u_{l_{k_2}} + \hat r_{l_{k_2}} < u_{l_{k_2+1}} + \hat r_{l_{k_2+1}} < u_{l_j} + \hat r_{l_j},
\eeaa
where the second inequality holds for $\e>0$ small enough. Now for the same  $\tilde \cC = \{1,\cds, \tilde k_*-1\}\cup\{l_{\tilde k_*}, \cds, l_{\h\k}\}$ with $|\tilde \cC|=\h\k$ as in Step $1.2.1$, we have
\beaa
&&\dis u_i + \hat r_i  \le u_{l_{\tilde k_*}} + \hat r_{l_{\tilde k_*}}  \le u_j + \hat r_j,\q \mbox{for all}~ i\in \tilde \cC,~ j\notin \tilde \cC;\\
 &&\dis \sum_{i\in \tilde\cC} [u_i + \hat r_i] = \sum_{i\in \tilde \cC} [u_i +  r_i] - [\h\k - \tilde k_* + 1]\e + \sum_{j= \tilde k_*}^{\h\k} \e = \sum_{i\in \tilde\cC} [u_i +  r_i]\ge 0.
\eeaa
Then by Lemma \ref{lem-r} (ii) we see that $\hat { r}\in \pmb \cS_R$. Moreover, note that
\beaa
\cT_{\hat { r}} - \cT_{ r} &=&{1\over 2}\Big[ |\hat r_1|-|r_1| + \sum_{j=\tilde k_*}^{k_2}[|\hat r_{l_j}|-|r_{l_j}|] + |\hat r_{l_{k_2+1}}|-  |r_{l_{k_2+1}}|\Big]\\
&=&{1\over 2}\Big[-[\h\k - \tilde k_* + 1]\e + \sum_{j=\tilde k_*}^{k_2}(-\e) + [k_2 - \h\k]\e\Big]  = -[\h\k - \tilde k_* + 1]\e <0,
\eeaa
where the last inequality is due to $\tilde k_* \le \h\k$ from Step $1.1.$  This contradicts the assumption that $r\in \pmb{\cS \cM}_R$ has the minimum total promises transfer, so equation
\reff{equilibrium2-claim2} holds true.

\medskip

{\bf Step 1.3.} We now collect all the results from the intermediate steps to show that $ r$ satisfies conditions \reff{rj<0}-\reff{cT* minimal} of Proposition~\ref{prop: Equilibria in presence of gain from trade2 Alternative}.

 Let $y_*$ denote the common value in \reff{equilibrium2-claim2}. Then $r_j = y_* - u_j\le 0$ for all $j\ge \tilde k_*$. On one hand, since $\tilde k_* \le k_* \le \h\k$ by Step $1.1$,
\beaa
0 &\le& \sum_{i\le \h\k} [u_i + r_i] = \sum_{i\le \h\k} u_i - \sum_{j>\h\k} r_j =  \sum_{i\le \h\k} u_i - \sum_{j>\h\k} [y_* - u_j] \\
&=& \sum_{i\in \dbI} u_i - (\k-1) y_* = (\k-1)(u_*-y_*).
\eeaa
Therefore, $y_* \le u_*$. On the other hand, by \reff{cTr+-},
\bea
\label{cTr>cT*}
\cT_{ r} = \sum_{j\ge \tilde k_*} (-r_j) \ge  \sum_{j\ge k_*} (-r_j) = \sum_{j\ge k_*} [u_j - y_*] \ge \sum_{j\ge k_*} [u_j - u_*] = \cT_*.
\eea
Since the net transfer profile $r$ minimizes the total promises transfer, and we already constructed a stable promises profile in Lemma~\ref{lem-example of promises}  with total promises transfer $\cT_*$, then we must have $\cT_{ r} = \cT_*$, and thus all the inequalities in \reff{cTr>cT*} are equalities. In particular, the second inequality in \reff{cTr>cT*} implies that $u_*= y_*$. Moreover, by \reff{rtildek*} and \reff{equilibrium2-claim2} we have $r_{\tilde k_*} < 0$, so that the first inequality in \reff{cTr>cT*} implies that $\tilde k_* = k_*$. Now it can be directly checked that the conditions \reff{rj<0}-\reff{cT* minimal} of Proposition~\ref{prop: Equilibria in presence of gain from trade2 Alternative} hold. This concludes the proof of the only if part of Proposition~\ref{prop: Equilibria in presence of gain from trade2 Alternative}.

\medskip

{\bf Step 2.} In this step, we show the if part: we fix $r\in \cP$
that satisfies conditions \reff{rj<0}-\reff{vi<vj} of Proposition~\ref{prop: Equilibria in presence of gain from trade2 Alternative} and show that $ r \in \pmb{\cS \cM}_R$ and that \reff{cT* minimal} holds.

To show that $\pmb r\in \cS_R$, set $\cC := \cC^S\cup \underline \cC^R$ in Lemma \ref{lem-r} (ii) so that $|\cC| = \h\k$. Using condition \reff{vi<vj} and observing that $u_j + r_j = u_*$ for all $j \ge k_*$ (condition \reff{ex post intensity}) shows that we have $u_i + r_i \le u_j+ r_j$ for all $i\in \cC$ and $j\notin \cC$. Moreover, since $ r \in \cP$ and $r_j =  -u_j + u_*$ for $j \geq \k_*$, we have
\beaa
\sum_{i\in \cC}(u_i+r_i) &=& \sum_{i\in \cC^S\cup \underline \cC^R} u_i - \sum_{j\in  \cC^R\backslash\underline \cC^R} r_j =  \sum_{i\in \cC^S\cup \underline \cC^R} u_i - \sum_{j\in  \cC^R\backslash\underline \cC^R} [-u_j +u_* ]\\
&=& \sum_{i\in \dbI} u_i - (\k-1) u_* =0.
\eeaa
Then by applying Lemma \ref{lem-r} (ii) we see that $ r\in \pmb \cS_R$.  Moreover,
\beaa
\cT_{ r} = \sum_{j\ge k_*} (-r_j) =  \sum_{j\ge k_*} [u_j - u_* ] = \cT_*.
\eeaa
By Proposition~\ref{prop-cEexistence} there exists $ r^* \in \pmb{\cS \cM}_R$. By Step $1$ (the only if direction), the net transfer profile $ r^*$  satisfies conditions \reff{rj<0}-\reff{cT* minimal}, in particular, $\cT_{ r^*} = \cT_*$, thus $\cT_*$ is the minimum total promises transfer for all $ r \in \pmb \cS_R$.  Finally, if $ r \in \pmb \cS_R$ satisfies  \reff{rj<0}-\reff{vi<vj},  we have $\cT_{ r} = \cT_*$, so $ r$ matches the minimum total promises transfer, and thus $ r\in \pmb{\cS \cM}_R$.
\qed

\newpage

\begin{center}
\subsubsection*{Supplemental Web Appendix B (SB): Reaching across the aisle transfer promises.}
\end{center}

\renewcommand{\theprop}{SB.\arabic{prop}} \setcounter{prop}{0}
\renewcommand\theequation{SB.\arabic{equation}} \setcounter{equation}{0}

In this appendix, we provide necessary and sufficient conditions under which the SM equilibrium transfer promises are of the reaching across the aisle type.  We state the proposition, discuss it and provide its proof.

\begin{prop} \label{prop: reaching across the aisle} [\textbf{Reaching across the aisle equilibria.}]
 When $\k \geq 2$,  all the SM equilibrium transfer promises are of the reaching across the aisle type if and only if one of the following conditions hold:

\begin{enumerate}

\item Reform supporters lack voting power to enact the reform, $|\cC^R|<\k$,  as in Proposition~\ref{prop: Equilibria with an intense minority}.
\item Reform supporters have enough voting power, $|\cC^R| \geq \k$, and $G^S \leq \D U_{\h \k}$ as in Proposition~\ref{prop: Equilibria in presence of gain from trade1}, and  one of the two following conditions are satisfied
\begin{enumerate}
\item $G^S = \Delta  U_{\hat\kappa}$, and the weakest reform supporters $\underline \cC^R=\{n+1, \cds, \h \k\}$  have equal   utilities: $u_{n+1} = \cds = u_{\hat\kappa}.$
\item $G^S < \Delta  U_{\hat\kappa}$ and the reform supporter $\h \k +1$ has a utility that is equal to that of the members $\{n+1, \cds, \h \k\}$:
    $u_{n+1} = \cds = u_{\hat\kappa} = u_{\hat\kappa +1}.$
\end{enumerate}

\item  Reform supporters have enough voting power, $|\cC^R| \geq \k$, and $ \Delta U_{\h \k} < G^S$ as in Proposition~\ref{prop: Equilibria in presence of gain from trade2 Alternative}, and  $u_{n+1} = \cds = u_{k_*-1} = u_*.$

\end{enumerate}
\end{prop}

Proposition~\ref{prop: reaching across the aisle} shows that
in the case of a majority coercion covered in Proposition~\ref{prop: Equilibria with an intense minority}, promises recipients are reform opponents in all SM equilibria. Despite their multiplicity, equilibrium transfers share the common feature of being of the reaching across the aisle type. By contrast, in the cases covered in Proposition~\ref{prop: Equilibria in presence of gain from trade1} and Proposition~\ref{prop: Equilibria in presence of gain from trade2 Alternative} where there are enough supporters to enact the reform to begin with ($ |\cC^R| \geq \k$), we show in  Proposition~\ref{prop: reaching across the aisle} that for the  SM equilibrium transfers to always be of the reaching across the aisle type, additional restrictions are required.  To rule out circle the wagon type transfers, we broadly need a stale distribution of intensities among a specific subset of reform supporters with weakest intensities.  In words, the proposition shows that when the weakest reform supporters derive uniform utility from the reform, then equilibrium requires that all promises recipients are reform opponents. We now give the proof of Proposition~\ref{prop: reaching across the aisle}.

\ms
\no {\bf Proof of Proposition~\ref{prop: reaching across the aisle}:} When $|\cC^R | < \k$, it is clear from Proposition~\ref{prop: Equilibria with an intense minority} that all promises recipients are  reform  opponents,
and thus statement $1$ in  Proposition~\ref{prop: reaching across the aisle} holds.

We now prove statement $2$ of  Proposition~\ref{prop: reaching across the aisle}. We first show that properties $2.a.$ and $2.b$ of Proposition~\ref{prop: reaching across the aisle} imply that any $ r\in \pmb{\cS \cM}_R$ belongs to the reaching across the aisle type.

Consider now the first subcase where $u_{n+1} = \cds =u_{\h k} $ and $G^S = \D U_{\h k}$. Note that
\beaa
u_j + r_j = \h \pi_j(r) \ge \h \pi_{\hat\k}(r) \ge u_{\hat\k},\q \forall j >\hat\k.
\eeaa
Then
\beaa
\cT_{ r} = \sum_{j>\hat\k} (-r_j) \le \sum_{j>\hat\k} (u_j - u_{\hat \k}) =  \Delta U_{\hat\kappa} = G^S.
\eeaa
Since $ r\in \pmb{\cS \cM}_R$, then $\cT_{ r} =G^S$ ( Proposition \ref{prop: Equilibria in presence of gain from trade1}, condition 3.), and thus equality holds above. This implies that $-r_j = u_j - u_{\hat \k}$, and thus $\h \pi_j(r) = u_{\hat\k}$ for all $j> \hat\k$. Note further that $u_i \le \h \pi_i(r) \le \h \pi_j(r) = u_{\hat\k}$ for all $n< i\le \hat\k < j$. By the assumption in this subcase, we see that $r_i =0$ for $n<i\le \hat\k$. Then the net transfer profile $ r$ is of the reaching across the aisle type.

Consider the second subcase where $u_{n+1} = \cds =u_{\h k} = u_{\h k+1}$.
By Proposition \ref{prop: Equilibria in presence of gain from trade1} Part 1, we have $r_i \ge 0$ for $ n< i\le \hat\k$ and $r_{\hat\k+1} \le 0$. Then $\h \pi_i(r) \ge u_i = u_{\hat\kappa + 1} \ge \h \pi_{\hat\k +1}(r)$. By Proposition \ref{prop: Equilibria in presence of gain from trade1} Part 2, we have $\h \pi_{\hat\k +1}(r) \geq \h \pi_i(r)$, and thus we must have $\h \pi_{\hat\k +1}(r) = \h \pi_i(r)$. Thus, $r_i = 0$ for $ n < i\le \hat\k$, and therefore $ r$ is of the reaching across the aisle type promise.

We next prove the only if part of statement 2 in Proposition~\ref{prop: Equilibria with an intense minority}. To do so, we assume that  both statement $2.a$ and statement $2.b$ are false, and construct an equilibrium promises profile $ r\in \pmb{\cS \cM}_R$ where some promises recipients are reform supporters. Note that, when   both statements $2.a$ and  $2.b$ are false, the assumption that $0<G^S\le \D U_{\hat\k}$ implies that one of the following two statements must be true:
\bea
\label{Case1}
&\dis u_{n+1} < u_{\hat\k}\q\mbox{ and }\q 0< G^S \le \D U_{\hat\k};\\
\label{Case2}
&\dis u_{n+1} = \cds = u_{\hat\k} < u_{\hat\k+1}\q\mbox{ and }\q 0< G^S < \D U_{\hat\k}.
\eea
Now let $0<\e< G^S$ and we modify the equilibrium promises profile in \reff{equilibrium1-example} as follows:
\bea
\label{equilibrium1-example-epsilon}
 \left.\ba{c}
 \dis r_i := -{G^S-\e\over U^S} u_i,~ i \in \cC^S;\q r_{n+1} := \e;\\
\dis  r_j := 0,~ n+1< j\le \hat\k;\q r_k := -{G^S\over {\D U_{\h\k}}}[u_k - u_{\h\k}],~ k\in \cC^R\backslash \underline \cC^R.
\ea\right.
\eea
It can be directly checked that $ r \in \cP$ and satisfies conditions \reff{equ: structure of r} and \reff{equilibrium1} from Proposition \ref{prop: Equilibria in presence of gain from trade1}, so to establish that $ r$ is an SM equilibrium, we only need to prove \reff{equ: ordering of intensities}.

In the sub-case \reff{Case1}, assume further that $\e < u_{\hat\k} - u_{n+1}$. Then one can check
\beaa
\h \pi_i(r) \le 0 \le \h \pi_j(r) \le u_{\hat \k} \le \h \pi_k(r),\q\mbox{for all}~ i \le n < j \le \hat\k < k.
\eeaa
In the sub-case \reff{Case2}, assume further that $\e < [1-{G^S\over \D U_{\hat\k}}][u_{\hat \k+1} - u_{\hat \k}]$. Then using $u_{n+1} = u_{\h \k}$ one can see that, for all $i \le n < j \le \hat\k < k$,
\beaa
\h \pi_i(r) \le 0 \le \h \pi_j(r) \le u_{\hat \k} +\e \le u_{\hat\k} + [1-{G^S\over \D U_{\hat\k}}][u_{\hat \k+1} - u_{\hat \k}] \le \h \pi_k(r).
\eeaa
To sum up, in all subcases, the promises profile $ r$ defined in \reff{equilibrium1-example-epsilon} also satisfies the condition \reff{equ: ordering of intensities} of Proposition \ref{prop: Equilibria in presence of gain from trade1} when $\e$ is small enough, and as a result, $ r\in \pmb{\cS \cM}_R$. We conclude then by observing that since $r_{n+1}>0$, the reform supporter $n+1$ is a promise recipient, and therefore the equilibrium $ r$ has some circle the wagon transfers.

We now prove statement $3$ in Proposition~\ref{prop: reaching across the aisle}.
We first show the if part. Fix an arbitrary $ r\in \pmb{\cS \cM}_R$.  By Proposition \ref{prop: Equilibria in presence of gain from trade2 Alternative}, we have $r_i \ge 0$ and $u_i \le \h \pi_i(r) \le u_*$ for $n+1 \le i\le \hat\k$. However, since we assume $u_{n+1} = \cds= u_{k_*-1} = u_*$ here, we must have $r_i =0$ for $n+1 \le i\le \hat\k$. Since $r_j = -(u_j - u_*)\ge 0$ for $j\ge k_*$, we see that the net transfer profile $ r$ is of the reaching across the aisle type.

We now prove the only if part. Recalling that the utilities $u_i$ are ordered, we assume $u_{n+1}< u_*$ and we shall construct an $ r\in \pmb{\cS \cM}_R$ which has some circle the wagon type transfers. Let $0<\e< \cT_*$ and  we modify the promises profile described in \reff{Example of equilibrium promises} as follows:
\bea \label{Example of equilibrium promises-epsilon}
\left.\ba{c}
\dis r_i := -{\cT_*-\e\over U^S} u_i,~ i \leq n;\qq r_{n+1} := \e;\\
\dis r_k := 0,~n+1< k < k_* ;\q r_j := -[u_k - u_*],~ j\geq k_*.
\ea\right.
\eea
Assume further that $\e < u_* - u_{n+1}$, then one can check that
\beaa
\h \pi_i(r) \le 0 \le \h \pi_j(r) \le u_*=\h \pi_k(r),\q\mbox{for all}~ i \le n < j \le \hat\k < k.
\eeaa
It can be checked that $ r \in \cP$ and satisfies all the other requirements in Proposition \ref{prop: Equilibria in presence of gain from trade2 Alternative}.
We conclude by observing that $r_{n+1}>0$, and therefore the SM equilibrium net transfer profile  $ r$ constructed in \reff{Example of equilibrium promises-epsilon} has some circle the wagon transfers.
\qed

 \end{document}